\documentclass[prodmode,acmcsur]{acmsmall}

\setcopyright{rightsretained}

\usepackage{dblfloatfix}
\usepackage[T1]{fontenc}
\usepackage{times}
\usepackage[cmex10]{amsmath}
\usepackage{amssymb} %
\usepackage[english]{babel}
\usepackage{url}
\usepackage[numbers,square,sort]{natbib}

\usepackage{dcolumn}
\usepackage{booktabs}
\usepackage[table,xcdraw]{xcolor}%
\usepackage{xparse}
\usepackage{float}
\usepackage[hidelinks]{hyperref} %
\usepackage{xspace}
\usepackage{listings}
\usepackage{subcaption}
\usepackage{fixltx2e}
\usepackage[figuresleft]{rotating}
\usepackage{afterpage}

\usepackage[super]{nth}

\usepackage{collcell}
\makeatletter
\newcolumntype{H}{>{\collectcell\@gobble}c<{\endcollectcell}@{}}
\makeatother

\usepackage[draft]{fixme}
\fxsetup{layout=inline,marginclue,theme=color}
\FXRegisterAuthor{jmn}{ajmn}{\color{red}jmn}
\FXRegisterAuthor{pl}{apl}{pl}
\FXRegisterAuthor{nb}{anb}{\color{blue}nb}
\FXRegisterAuthor{sb}{asb}{\color{magenta}sbr}
\FXRegisterAuthor{sc}{asc}{\color{orange}sc}
\FXRegisterAuthor{mp}{amp}{\color{teal}mp}
\FXRegisterAuthor{mf}{amf}{\color{pink}mf}

\usepackage{microtype}
\usepackage{tikz}
\usepackage{balance}
\usetikzlibrary{arrows,chains,calc,matrix,positioning,scopes,shapes,decorations.pathreplacing,shapes.arrows,shapes.multipart,patterns,automata}

\newcommand{\SpecCPU}{SPEC~CPU2006\xspace}
\newcommand{\CFGuard}{CFGuard\xspace} %
\newcommand{\MCFI}{MCFI\xspace} %
\newcommand{\piCFI}{$\pi$CFI\xspace} %
\newcommand{\IFCC}{IFCC\xspace} %
\newcommand{\VTV}{VTV\xspace} %
\newcommand{\CryptoCFI}{C-CFI\xspace}
\newcommand{\LLVMCFI}{LLVM-CFI\xspace} %
\newcommand{\CCFIR}{CCFIR\xspace} %
\newcommand{\SD}{SafeDispatch\xspace} %
\newcommand{\LD}{Lockdown\xspace} %
\newcommand{\OCFI}{OCFI\xspace} %
\newcommand{\ROPecker}{ROPecker\xspace} %
\newcommand{\KBouncer}{kBouncer\xspace} %
\newcommand{\CFIMon}{CFIMon\xspace} %
\newcommand{\ROPGuard}{ROPGuard\xspace} %
\newcommand{\PathArmor}{PathArmor\xspace} %
\newcommand{\VTI}{VTI\xspace} %
\newcommand{\binCFI}{bin-CFI\xspace}

\newcommand{\gray}{\rowcolor[HTML]{E3E3E3} }

\usepackage{relsize}
\def\ifmonospace{\ifdim\fontdimen3\font=0pt }
\def\C++{%
\ifmonospace%
    \C++%
\else%
    C\kern-.1167em\raise.30ex\hbox{\smaller{++}}%
\fi%
\spacefactor1000 }
\def\lib\C++{%
\ifmonospace%
    lib\C++%
\else%
    libc\kern-.1167em\raise.20ex\hbox{\smaller{++}}%
\fi%
\spacefactor1000 }

\NewDocumentCommand{\rot}{O{45} O{1em} m}{\makebox[#2][l]{\rotatebox{#1}{#3}}}%

\newcommand{\UNITS}{11} %

\newdimen\R %
\newdimen\L %
\tikzset{%
    rotated/.style={rotate=45},
    comp25/.style={scale=.8},
    comp4/.style={scale=.9,rotate=30},
    comp3/.style={scale=.9,},
}

\newdimen\LSAPF %
\newcommand{\QDIMENSIONS}{5} %
\newcommand{\QA}{360/\QDIMENSIONS} %
\newcommand{\drawQuantSpider}{%
  \tikzstyle{source}=[color=green,line width=1.5pt,opacity=0.5]
  \tikzstyle{binary}=[color=blue,line width=1.5pt,opacity=0.5]
  \tikzstyle{runtime}=[color=red,line width=1.5pt,opacity=0.5]
  \tikzstyle{hardware}=[color=red,line width=1.5pt,opacity=0.5]
  \path (0:0cm) coordinate (O); %

  \foreach \X in {1,...,\QDIMENSIONS}{
    \draw (\X*\QA:0) -- (\X*\QA:\R);
  }

  \foreach \Y in {0,...,\UNITS}{
    \foreach \X in {1,...,\QDIMENSIONS}{
      \path (\X*\QA:\Y*\R/\UNITS) coordinate (D\X-\Y);
      \fill (D\X-\Y) circle (.6pt);
    }
  }
  \path (1*\QA:\L) node (L1) {\tiny CF};
  \path (2*\QA:\LSAPF) node (L3) {\tiny SAP.F};
  \path (3*\QA:\L) node (L4) {\tiny SAP.B};
  \path (4*\QA:\L) node (L5) {\tiny RP}; %
  \path (5*\QA:\L) node (L6) {\tiny Q};  %
}

\newcommand{\FDIMENSIONS}{4} %
\newcommand{\FA}{360/\FDIMENSIONS} %
\newcommand{\drawRotatedSpider}{%
  \tikzstyle{source}=[color=green,line width=1.5pt,opacity=0.5]
  \tikzstyle{binary}=[color=blue,line width=1.5pt,opacity=0.5]
  \tikzstyle{runtime}=[color=red,line width=1.5pt,opacity=0.5]
  \tikzstyle{hardware}=[color=red,line width=1.5pt,opacity=0.5]
  \path (0:0cm) coordinate (O); %

  \foreach \X in {1,...,5}{
    \draw (\X*\FA:0) -- (\X*\FA:\R);
  }
  \foreach \Y in {0,...,\UNITS}{
    \foreach \X in {1,...,\FDIMENSIONS}{
      \path (\X*\FA:\Y*\R/\UNITS) coordinate (D\X-\Y);
      \fill (D\X-\Y) circle (.6pt);
    }
  }
  \path (1*\FA:\L) node (L1) {\tiny CF};
  \path (2*\FA:\L) node (L2) {\tiny SAP.F};
  \path (3*\FA:\L) node (L3) {\tiny SAP.B};
  \path (4*\FA:\L) node (L4) {\tiny RP}; %
}

\newcommand{\ZDIMENSIONS}{3} %
\newcommand{\ZA}{360/\ZDIMENSIONS} %
\newcommand{\drawFwdOnlySpider}{%
  \tikzstyle{source}=[color=green,line width=1.5pt,opacity=0.5]
  \tikzstyle{binary}=[color=blue,line width=1.5pt,opacity=0.5]
  \tikzstyle{runtime}=[color=red,line width=1.5pt,opacity=0.5]
  \tikzstyle{hardware}=[color=red,line width=1.5pt,opacity=0.5]
  \path (0:0cm) coordinate (O); %

  \foreach \X in {1,...,\ZDIMENSIONS}{
    \draw (\X*\ZA:0) -- (\X*\ZA:\R);
  }

  \foreach \Y in {0,...,\UNITS}{
    \foreach \X in {1,...,\ZDIMENSIONS}{
      \path (\X*\ZA:\Y*\R/\UNITS) coordinate (D\X-\Y);
      \fill (D\X-\Y) circle (.6pt);
    }
  }
  \path (1*\ZA:\L) node (L1) {\tiny CF};
  \path (2*\ZA:\L) node (L3) {\tiny SAP.F};
  \path (3*\ZA:\L) node (L5) {\tiny RP}; %
}

\newcommand{\drawQFwdOnlySpider}{%
  \tikzstyle{source}=[color=green,line width=1.5pt,opacity=0.5]
  \tikzstyle{binary}=[color=blue,line width=1.5pt,opacity=0.5]
  \tikzstyle{runtime}=[color=red,line width=1.5pt,opacity=0.5]
  \tikzstyle{hardware}=[color=red,line width=1.5pt,opacity=0.5]
  \path (0:0cm) coordinate (O); %

  \foreach \X in {1,...,\ZDIMENSIONS}{
    \draw (\X*\ZA:0) -- (\X*\ZA:\R);
  }

  \foreach \Y in {0,...,\UNITS}{
    \foreach \X in {1,...,\ZDIMENSIONS}{
      \path (\X*\ZA:\Y*\R/\UNITS) coordinate (D\X-\Y);
      \fill (D\X-\Y) circle (.6pt);
    }
  }
  \path (1*\ZA:\L) node (L1) {\tiny CF};
  \path (2*\ZA:\L) node (L3) {\tiny SAP.F};
  \path (3*\ZA:\L) node (L5) {\tiny Q}; %
}

\newcommand{\drawQRotatedSpider}{%
  \tikzstyle{source}=[color=green,line width=1.5pt,opacity=0.5]
  \tikzstyle{binary}=[color=blue,line width=1.5pt,opacity=0.5]
  \tikzstyle{runtime}=[color=red,line width=1.5pt,opacity=0.5]
  \tikzstyle{hardware}=[color=red,line width=1.5pt,opacity=0.5]
  \path (0:0cm) coordinate (O); %

  \foreach \X in {1,...,5}{
    \draw (\X*\FA:0) -- (\X*\FA:\R);
  }
  \foreach \Y in {0,...,\UNITS}{
    \foreach \X in {1,...,\FDIMENSIONS}{
      \path (\X*\FA:\Y*\R/\UNITS) coordinate (D\X-\Y);
      \fill (D\X-\Y) circle (.6pt);
    }
  }
  \path (1*\FA:\L) node (L1) {\tiny CF};
  \path (2*\FA:\L) node (L2) {\tiny SAP.F};
  \path (3*\FA:\L) node (L3) {\tiny RP};
  \path (4*\FA:\L) node (L4) {\tiny Q}; %
}

\title{\LARGE Control-Flow Integrity: Precision, Security, and Performance}

\author{
Nathan Burow
\affil{Purdue University}
Scott A. Carr
\affil{Purdue University}
Joseph Nash
\affil{University of California, Irvine}
Per Larsen
\affil{University of California, Irvine}
Michael Franz
\affil{University of California, Irvine}
Stefan Brunthaler
\affil{Paderborn University \& SBA Research}
Mathias Payer
\affil{Purdue University}
}

\begin{document}
\renewcommand{\sectionautorefname}{Section}
\renewcommand{\subsectionautorefname}{Section}
\renewcommand{\subsubsectionautorefname}{Section}
\renewcommand{\figureautorefname}{Figure}
\renewcommand{\tableautorefname}{Table}

\begin{abstract}

Memory corruption errors in C/\C++ programs remain the most common source of
security vulnerabilities in today's systems. Control-flow hijacking attacks
exploit memory corruption vulnerabilities to divert program execution away from
the intended control flow. Researchers have spent more than a decade studying
and refining defenses based on Control-Flow Integrity (CFI), and this technique
is now integrated into several production compilers. However, so far no study
has systematically compared the various proposed CFI mechanisms, nor is there
any protocol on how to compare such mechanisms.

We compare a broad range of CFI mechanisms using a unified nomenclature based on
(i)~a qualitative discussion of the conceptual security guarantees, (ii)~a
quantitative security evaluation, and (iii)~an empirical evaluation of their
performance in the same test environment. For each mechanism, we evaluate
(i)~protected types of control-flow transfers, (ii)~the precision of the
protection for forward and backward edges.  For open-source compiler-based
implementations, we additionally evaluate (iii)~the generated equivalence
classes and target sets, and (iv)~the runtime performance.
\end{abstract}



\acmformat{Nathan Burow, Scott A. Carr, Joseph Nash, Per Larsen, Michael Franz,
  Stefan Brunthaler, and Mathias Payer. 2016. Control-Flow Integrity: Precision,
Security, and Performance.}

\begin{bottomstuff}
This material is based upon work supported, in part, by the National Science
Foundation under Grant No. CNS-1464155, CNS-1513783, CNS-1657711, 
CNS-1513837, CNS-1619211 and
IIP-1520552, and by the
Defense Advanced Research Projects Agency (DARPA) under contracts
FA8750-15-C-0124, FA8750-15-C-0085, and
FA8750-10-C-0237,
and by COMET K1 of the Austrian Research Promotion Agency (FFG),
as well as gifts from Intel, Mozilla, Oracle,
and Qualcomm.
Any opinions, findings, and conclusions or recommendations expressed in this
material are those of the authors and do not necessarily reflect the views of
the Defense Advanced Research Projects Agency (DARPA), its Contracting Agents,
the National Science Foundation, or any other agency of the U.S. Government.

Author's addresses: N. Burow, S. A. Carr, and M. Payer, Purdue University,
Department of Computer Science, \{burow, carr27, mpayer\}@purdue.edu.
J. Nash, P. Larsen, and M. Franz, University of California-Irvine, Department of
Computer Science, \{jmnash, perl, franz\}@uci.edu.
S. Brunthaler, Paderborn University, Department of Computer Science,
s.brunthaler@upb.de.
\end{bottomstuff}

\maketitle

\section{Introduction}\label{sec:intro}
Systems programming languages such as C and \C++ give programmers a high degree
of freedom to optimize and control how their code uses available resources.
While this facilitates the construction of highly efficient programs, requiring
the programmer to manually manage memory and observe typing rules leads to
security vulnerabilities in practice.   Memory corruptions, such as buffer
overflows, are routinely exploited by attackers.  Despite significant research
into exploit mitigations, very few of these mitigations have entered
practice~\cite{szekeres.etal+13}. The combination of three such defenses,
(i)~Address Space Layout Randomization (ASLR)~\cite{ASLR}, (ii)~stack
canaries~\cite{execshield}, and (iii)~Data Execution Prevention (DEP)~\cite{DEP}
protects against \emph{code-injection attacks,} but are unable to fully prevent
\emph{code-reuse attacks}. Modern exploits use Return-Oriented Programming (ROP)
or variants thereof to bypass currently deployed defenses and divert the control
flow to a malicious payload. Common objectives of such payloads include
arbitrary code execution, privilege escalation, and exfiltration of sensitive
information.

The goal of Control-Flow Integrity (CFI)~\cite{abadi.etal+05} is to restrict the
set of possible control-flow transfers to those that are strictly required for
correct program execution. This prevents code-reuse techniques such as ROP from
working because they would cause the program to execute control-flow transfers
which are illegal under CFI. Conceptually, most CFI mechanisms follow a
two-phase process. An \emph{analysis} phase constructs the Control-Flow Graph
(CFG) which approximates the set of legitimate control-flow transfers. This CFG
is then used at runtime by an \emph{enforcement} component to ensure that all
executed branches correspond to edges in the CFG.

During the analysis phase, the CFG is computed by analyzing either the source
code or binary of a given program. In either case, the limitations of static
program analysis lead to an over-approximation of the control-flow transfers
that can actually take place at runtime. This over-approximation limits the
security of the enforced CFI policy because some non-essential edges are
included in the CFG.

The enforcement phase ensures that control-flow transfers which are potentially
controlled by an attacker, i.e., those whose targets are computed at runtime,
such as indirect branches and return instructions, correspond to
edges in the CFG produced by the analysis phase.  These targets are commonly
divided into forward edges such as indirect calls, and backward edges like
return instructions (so called because they return control back to the calling
function). All CFI mechanisms protect forward edges, but some do not handle
backward edges. Assuming code is static and immutable 
\footnote{DEP marks code pages as executable and readable by default.
Programs may subsequently change permissions to make code pages writable using
platform-specific APIs such as {\tt mprotect}. Mitigations such as PaX
\textsc{MPROTECT}, SELinux~\cite{mccarty+04}, and the {\tt
ProcessDynamicCodePolicy} Windows API restrict how page permissions can be
changed to prevent code injection and modification.},
CFI can be enforced by instrumenting existing indirect control-flow transfers
at compile time through a modified compiler, ahead of time through static binary
rewriting, or during execution through dynamic binary translation.  The types of
indirect transfers that are
subject to such validation and the number of valid targets per branch varies
greatly between different CFI defenses.  These differences have a major impact
on the security and performance of the CFI mechanism.

CFI does not seek to prevent memory corruption, which is the root cause of most
vulnerabilities in C and \C++ code.  While mechanisms that enforce
spatial~\cite{softbound} and temporal~\cite{nagarakatte2010ismm} memory safety
eliminate memory corruption (and thereby control-flow hijacking attacks),
existing mechanisms are considered prohibitively expensive.  In contrast, CFI
defenses offer reasonably low overheads while making it substantially harder for
attackers to gain arbitrary code execution in vulnerable programs. Moreover, CFI
requires few changes to existing source code which allows complex software to be
protected in a mostly automatic fashion.
While the idea of restricting branch instructions based on target sets predates
CFI~\cite{kiriansky02sec, kiriansky03thesis, PaX},
Abadi et al.'s seminal paper~\cite{abadi.etal+05} was the first formal
description of CFI with an accompanying implementation.  Since this paper was
published over a decade ago, the research community has proposed a large number
of variations of the original idea. More recently, CFI implementations have been
integrated into production-quality compilers, tools, and operating systems.

Current CFI mechanisms can be compared along two major axes: performance and
security. In the scientific literature, performance overhead is usually measured
through the \SpecCPU benchmarks. Unfortunately, sometimes only a subset of the
benchmarks is used for evaluation. To evaluate security, many authors have used
the Average Indirect target Reduction (AIR)~\cite{cfi-cots} metric that counts
the overall reduction of targets for any indirect control-flow transfer.

Current evaluation techniques do not adequately distinguish among CFI mechanisms
along these axes.  Performance measurements are all in the same range, between
0\% and 20\% across different benchmarks with only slight variations for the
same benchmark. Since the benchmarks are evaluated on different machines with
different compilers and software versions, these numbers are close to the margin
of measurement error. On the security axis, AIR is not a desirable metric for
two reasons.  First, all CFI mechanisms report similar AIR numbers (a $>99\%$
reduction of branch targets), which makes AIR unfit to compare individual CFI
mechanisms against each other.  Second, even a large reduction of targets often
leaves enough targets for an attacker to achieve the desired
goals~\cite{out-of-control, ROP-CFI:Davi, ROP-CFI:Wagner}, making AIR unable to
evaluate security of CFI mechanisms on an absolute scale.  %
We systematize the different CFI mechanisms (where ``mechanism'' captures both
the analysis and enforcement aspects of an implementation) and compare
them against metrics for security and performance.  By introducing metrics for
these areas, our analysis allows the objective comparison of different CFI
mechanisms both on an absolute level and relatively against other mechanisms.
This in turn allows potential users to assess the trade-offs of individual CFI
mechanisms and choose the one that is best suited to their use case.  Further,
our systematization provides a more meaningful way to classify CFI
mechanism than the ill-defined and inconsistently used ``coarse'' and
``fine'' grained classification.

To evaluate the security of CFI mechanisms we follow a \emph{comprehensive}
approach, classifying them according to a \emph{qualitative} and a
\emph{quantitative} analysis.  In the qualitative security discussion we compare
the strengths of individual solutions on a conceptual level by evaluating the
CFI policy of each mechanism along several axes: (i)
precision in the forward direction, (ii) precision in the backward direction,
(iii) supported control-flow transfer types according to the source
programming language, and (iv) reported performance.  In the
quantitative evaluation, we measure the target sets generated by each
CFI mechanism for the \SpecCPU benchmarks. The precision and security
guarantees of a CFI mechanism depend on the \emph{precision} of
target sets used at runtime, i.e., across all control-flow transfers,
how many superfluous targets are reachable through an individual
control-flow transfer. We compute these target sets for all available
CFI mechanisms and compare the ranked sizes of the sets against each
other. This methodology lets us compare the actual sets used for the
integrity checks of one mechanism against other mechanisms. In
addition, we collect all indirect control-flow targets used for the
individual \SpecCPU benchmarks and use these sets as a lower bound on
the set of required targets.  We use this lower bound to compute how
close a mechanism is to an \emph{ideal} CFI mechanism.  An ideal CFI
mechanism is one where the enforced CFG's edges exactly correspond to the
executed branches.

As a second metric, we evaluate the performance impact of open-sourced,
compiler-based CFI mechanisms. In their corresponding publications, each
mechanism was evaluated on different hardware, different libraries, and
different operating systems, using either the full or a partial set of \SpecCPU
benchmarks. We cannot port all evaluated CFI mechanisms to the same baseline
compiler. Therefore, we measure the overhead of each mechanism relative to the
compiler it was integrated into. This apples-to-apples comparison
highlights which \SpecCPU benchmarks are most useful when evaluating CFI.
The paper is structured as follows.  We first give a detailed background of the
theory underlying the analysis phase of CFI mechanisms.  This allows us to then
qualitatively compare the different mechanisms on the precision of their
analysis.  We then quantify this comparison with a novel metric.  This is
followed by our performance results for the different implementation.  Finally,
we highlight best practices and future research directions for the CFI community
identified during our evaluation of the different mechanisms, and conclude.

Overall, we present the following contributions:

\begin{enumerate}
\item a systematization of CFI mechanisms with a focus on discussing
the major different CFI mechanisms and their respective trade-offs,

\item a taxonomy for classifying the underlying analysis of a CFI mechanism,

\item presentation of both a qualitative and quantitative security metric
and the evaluation of existing CFI mechanisms along these metrics, and

\item a detailed performance study of existing CFI mechanisms.

\end{enumerate}

\section{Foundational Concepts}\label{s:foundations}
We first introduce CFI and discuss the two components
of most CFI mechanisms: (i) the \emph{analysis} that defines the CFG (which
inherently limits the precision that can be achieved) and (ii) the runtime
instrumentation that \emph{enforces} the generated CFG. Secondly, we classify
and systematize different types of control-flow transfers and how they are used
in programming languages.  Finally, we briefly discuss the CFG precision
achievable with different types of static analysis.  For those interested, a
more comprehensive overview of static analysis techniques is available in
\autoref{app:sa-prior-work}.

\subsection{Control-Flow Integrity}\label{ss:cfi}
CFI is a policy that restricts the execution flow of a program at runtime to
a predetermined CFG by validating indirect control-flow transfers. On the
machine level, indirect control-flow transfers may target any executable address
of mapped memory, but in the source language (C, \C++, or Objective-C) the
targets are restricted to valid language constructs such as functions, methods
and switch statement cases. Since the aforementioned languages rely on manual
memory management, it is left to the programmer to ensure that non-control data
accesses do not interfere with accesses to control data such that programs
execute legitimate control flows. Absent any security policy, an attacker can
therefore exploit memory corruption to redirect the control-flow to an arbitrary
memory location, which is called control-flow hijacking.
CFI closes the gap between machine and source code semantics by restricting the
allowed control-flow transfers to a smaller set of target locations. This
smaller set is determined per indirect control-flow location.  Note that
languages providing complete memory and type safety generally do not need to be
protected by CFI. However, many of these ``safe'' languages rely on virtual
machines and libraries written in C or \C++ that will benefit from CFI
protection.

\begin{figure}[t]
  \centering
  \begin{lstlisting}[language=C++,numbers=left,mathescape=true]
  void foo(int a){
      return;
  }
  void bar(int a){
      return;
  }
  void baz(void){
      int a = input();
      void (*fptr)(int);
      if(a){
        fptr = foo;
        fptr();
      } else {
        fptr = bar;
        fptr();
      }
  }
  \end{lstlisting}
  \caption{Simplified example of over approximation in static analysis.}
  \label{fig:sa-example}
\end{figure}

Most CFI mechanisms determine the set of valid targets for each indirect control-flow
transfer by computing the CFG of the program.  The security guarantees of a CFI
mechanism depend on the precision of the CFG it constructs.  The CFG cannot be
perfectly precise for non-trivial programs.  Because the CFG is statically
determined, there is always some over-approximation due to imprecision of the
static analysis.  An equivalence class is the set of valid targets for a given
indirect control-flow transfer.  Throughout the following, we reference
\autoref{fig:sa-example}.  Assuming an analysis based on function types or a
flow-insensitive analysis, both
\texttt{foo()} and \texttt{bar()} end up in the same equivalence class.  Thus,
at line 12 and line 15 either function can be called.  However, from the source
code we can tell that at line 12 only \texttt{foo()} should be called, and at
line 15 only \texttt{bar()} should be called. While this specific problem can be
addressed with a flow-sensitive analysis, all known static program analysis
techniques are subject to some over-approximation (see
\autoref{app:sa-prior-work}).

Once the CFI mechanism has computed an approximate CFG, it has to enforce its
security policy. We first note that CFI does not have to enforce constraints for
control-flows due to direct branches because their targets are immune to memory
corruption thanks to DEP. Instead, it focuses on attacker-corruptible branches
such as indirect calls, jumps, and returns. In particular, it must protect
control-flow transfers that allow runtime-dependent, targets such as
\texttt{void (*fptr)(int)} in \autoref{fig:sa-example}. These targets are stored
in either a register or a memory location depending on the compiler and the
exact source code. The indirection such targets provide allows flexibility as,
e.g., the target of a function may depend on a call-back that is passed from
another module. Another example of indirect control-flow transfers is return
instructions that read the return address from the stack. Without such an
indirection, a function would have to explicitly enumerate all possible callers
and check to which location to return to based on an explicit comparison.

For indirect call sites, the CFI enforcement component validates target
addresses before they are used in an indirect control-flow transfer. This
approach detects code pointers (including return addresses) that were modified
by an attacker -- if the attacker's chosen target is not a member of the
statically determined set.

\subsection{Classification of Control-Flow Transfers}
\label{ss:cflow-classification}

Control-flow transfers can broadly be separated into two categories: (i)
\emph{forward} and (ii) \emph{backward}.  Forward control-flow transfers are
those that move control to a new location inside a program.
When a program returns control to a prior location, we call this a backward
control-flow\footnote{Note the ambiguity of a backward edge in machine code
(i.e., a backward jump to an earlier memory location) which is different from a
backward control-flow transfer as used in CFI.}.

A CPU's instruction-set architecture (ISA) usually offers two forward
control-flow transfer instructions: call and jump.  Both of these are either
direct or indirect, resulting in four different types of forward control-flow:

\begin{itemize}
\item \emph{direct jump}:
  is a jump to a constant, statically determined target address.
  Most local control-flow, such as loops or if-then-else cascaded statements,
  use direct jumps to manage control.

\item \emph{direct call}:
  is a call to a constant, statically determined target address.
  Static function calls, for example, use direct call instructions.

\item \emph{indirect jump}:
  is a jump to a computed, i.e., dynamically determined target address.
  Examples for indirect jumps are switch-case statements using a dispatch
  table, Procedure Linkage Tables (PLT),
  as well as the threaded code interpreter dispatch optimization~\cite{bell+73,kogge+82,debaere.campenhout+90}.

\item \emph{indirect call}:
  is a call to a computed, i.e., dynamically determined target address.
  The following three examples are relevant in practice:

    \textbf{Function pointers}
    are often used to emulate object-oriented method dispatch in classical
    record data structures, such as C \texttt{struct}s,
    or for passing callbacks to other functions.

    \textbf{vtable dispatch}
    is the preferred way to implement dynamic dispatch to \C++ methods.
    A \C++ object keeps a pointer to its \emph{vtable}, a table containing
    pointers to all virtual methods of its dynamic type.  A method call,
    therefore, requires (i) dereferencing the vtable pointer, (ii) computing
    table index using the method offset determined by the object's static type,
    and (iii) an indirect call instruction to the table entry referenced
    in the previous step.  In the presence of multiple inheritance, or multiple
    dispatch, dynamic dispatch is slightly more complicated. %
    \textbf{Smalltalk-style \texttt{send}-method dispatch} that requires a dynamic
    type look-up.
    Such a dynamic dispatch using a \texttt{send}-method in Smalltak,
    Objective-C, or JavaScript
    requires walking the class hierarchy (or the prototype chain in
    JavaScript) and selecting the first method with a matching identifier.
    This procedure is required for all method calls and therefore
    impacts performance negatively. Note that, e.g., Objective-C uses a lookup
    cache to reduce the overhead.
\end{itemize}

We note that jump instructions can also be either conditional or unconditional.
For the purposes of this paper this distinction is irrelevant.

All common ISAs support backward and forward indirect control-flow transfers.
For example, the x86 ISA supports backward control-flow transfers using just
one instruction: return, or just \texttt{ret}.  A return instruction is the
symmetric counterpart of a call instruction, and a compiler emits function
prologues and epilogues to form such pairs.  A call instruction pushes the
address of the immediately following instruction onto the native machine stack.
A return instruction pops the address off the native machine stack and updates
the CPU's instruction pointer to point to this address.  Notice that a return
instruction is conceptually similar to an indirect jump instruction, since the
return address is unknown a priori.  Furthermore, compilers are emitting
call-return pairs by \emph{convention} that hardware usually does not enforce.
By modifying return addresses on the stack, an attacker can ``return'' to
all addresses in a program, the foundation of return-oriented
programming~\cite{ret2libc,Shacham10,Roemer2011}.

Control-flow transfers can become more complicated in the presence of
exceptions.  Exception handling complicates control-flows locally, i.e., within
a function, for example by moving control from a try-block into a catch-block.
Global exception-triggered control-flow manipulation, i.e., interprocedural
control-flows, require unwinding stack frames on the current stack until a
matching exception handler is found.

Other control-flow related issues that CFI mechanisms should (but not always do)
address are: (i) separate compilation, (ii) dynamic linking, and (iii) compiling
libraries.  These present challenges because the entire CFG may not be known at
compile time.  This problem can be solved by relying on LTO, or dynamically
constructing the combined CFG.  Finally, keep in mind that, in general, not all
control-flow transfers can be recovered from a binary.

Summing up, our classification scheme of control-flow transfers is as follows:
\begin{itemize}
\itemsep-0pt
\item \textbf{CF.1}: backward control-flow,
\item \textbf{CF.2}: forward control-flow using direct jumps,
\item \textbf{CF.3}: forward control-flow using direct calls,
\item \textbf{CF.4}: forward control-flow using indirect jumps,
\item \textbf{CF.5}: forward control-flow using indirect calls supporting function pointers,
\item \textbf{CF.6}: forward control-flow using indirect calls supporting vtables,
\item \textbf{CF.7}: forward control-flow using indirect calls supporting
  Smalltalk-style method dispatch,
\item \textbf{CF.8}: complex control-flow to support exception handling,
\item \textbf{CF.9}: control-flow supporting language features such as dynamic linking, separate compilation, etc.
\end{itemize}

According to this classification, the C programming language uses control-flow
transfers 1--5, 8 (for setjmp/longjmp) and 9, whereas the \C++ programming
language allows all control-flow transfers except no.~7.

\subsection{Classification of Static Analysis Precision}\label{sss:sap}

As we saw in \autoref{ss:cfi}, the security guarantees of a CFI mechanism
ultimately depend on the precision of the CFG that it computes.
This precision is in turn determined by the type of static analysis used.
For the purposes of this paper, the following classification
summarizes prior work to determine forward control-flow transfer analysis
precision (see \autoref{app:sa-prior-work} for full details).  In order of
increasing static analysis precision (SAP), our classifications are:

\begin{itemize}
\itemsep-0pt
\item \textbf{SAP.F.0}: No forward branch validation
\item \textbf{SAP.F.1a}: ad-hoc algorithms and heuristics
\item \textbf{SAP.F.1b}: context- and flow-insensitive analysis
\item \textbf{SAP.F.1c}: labeling equivalence classes
\item \textbf{SAP.F.2}: class-hierarchy analysis
\item \textbf{SAP.F.3}: rapid-type analysis
\item \textbf{SAP.F.4a}: flow-sensitive analysis
\item \textbf{SAP.F.4b}: context-sensitive analysis
\item \textbf{SAP.F.5}: context- and flow-sensitive analysis
\item \textbf{SAP.F.6}: dynamic analysis (optimistic)
\end{itemize}

The following classification
summarizes prior work to determine backward control-flow transfer analysis
precision:
\begin{itemize}
\itemsep-0pt
\item \textbf{SAP.B.0}: No backward branch validation
\item \textbf{SAP.B.1}: Labeling equivalence classes
\item \textbf{SAP.B.2}: Shadow stack
\end{itemize}

Note that there is well established and vast prior work in static analysis that
goes well beyond the scope of this paper~\cite{nielson09springer}.  The goal of
our systematization is merely to summarize the most relevant aspects and use
them to shed more light on the precision aspects of CFI.

\subsection{Nomenclature and Taxonomy}
Prior work on CFI usually classifies mechanisms into fine-grained and coarse-grained.
Over time, however, these terms have been used to describe different systems
with varying granularity and have, therefore, become overloaded and imprecise.
In addition, prior work only uses a rough separation into forward and backward
control-flow transfers without considering sub types or precision.  We hope that
the classifications here will allow a more precise and consistent definition of
the precision of CFI mechanisms underlying analysis, and will encourage the CFI
community to use the most precise techniques available from the static analysis
literature.

\section{Security}\label{sec:security}

In this section we present a security analysis of existing CFI implementations.
Drawing on the foundational knowledge in \autoref{s:foundations}, we present a
qualitative analysis of the theoretical security of different CFI mechanisms
based on the policies that they implement.  We then give a quantitative
evaluation of a selection of CFI implementations.  Finally, we survey previous
security evaluations and known attacks against CFI.

\afterpage{%
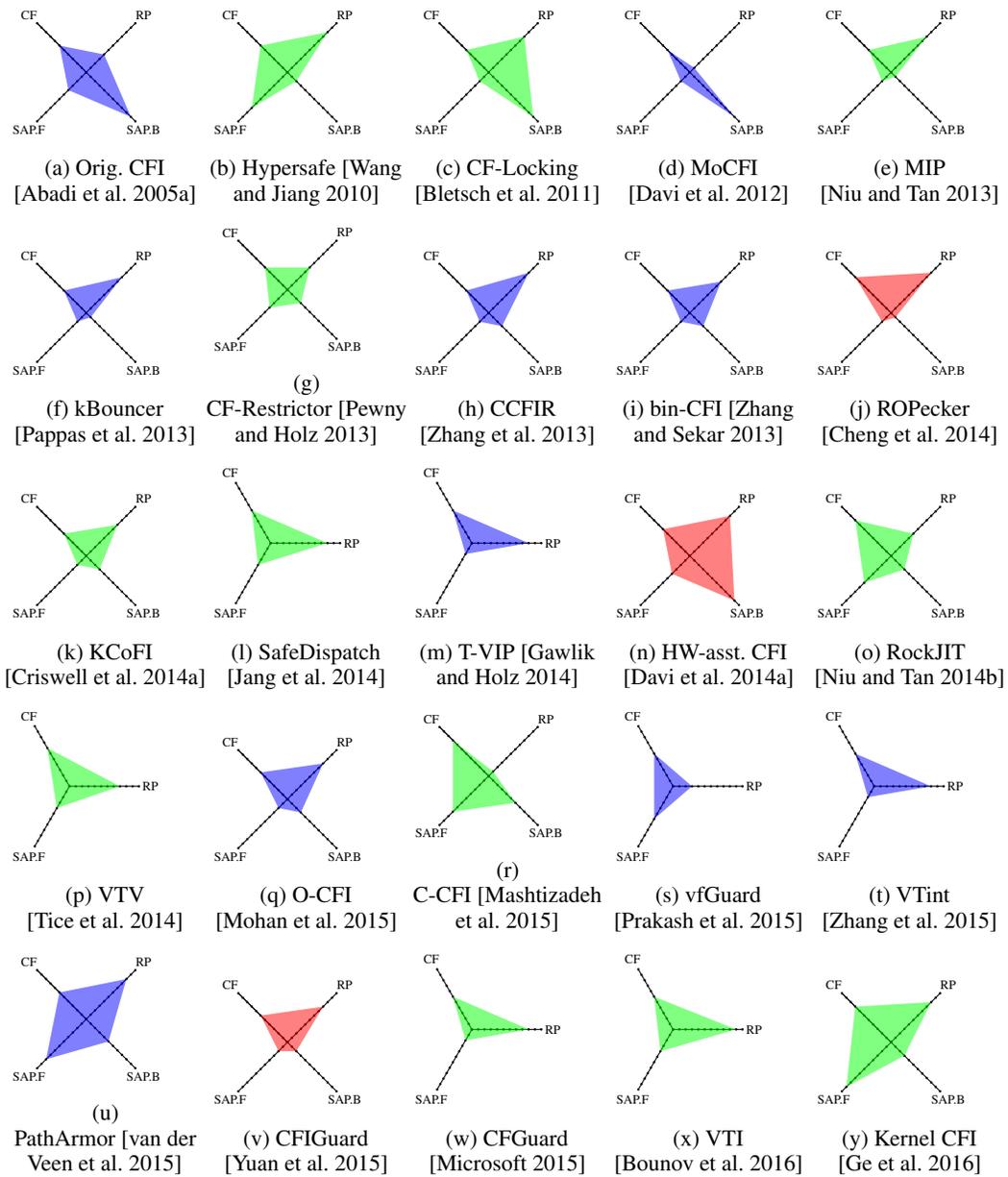
\begin{figure}
\captionsetup[subfigure]{justification=centering}
 \centering
  \begin{subfigure}[b]{.2\textwidth}
    \begin{tikzpicture}[comp25,rotated]
      \drawRotatedSpider{}
      \fill [binary]
      (D1-6) --                   %
      (D2-4) --
      (D3-10) --
      (D4-4) -- cycle;
    \end{tikzpicture}
    \caption{Orig. CFI\\ \cite{abadi.etal+05}%
     }\label{fig:spider:orig-cfi}
  \end{subfigure}%
  \begin{subfigure}[b]{.2\textwidth}
    \begin{tikzpicture}[comp25,rotated]
      \drawRotatedSpider{}
      \fill [source]
      (D1-6) --                   %
      (D2-8) --                   %
      (D3-2) --                  %
      (D4-9) -- cycle;            %
    \end{tikzpicture}
    \caption{Hypersafe \cite{wang10oakland}%
     }\label{fig:spider:hypersafe}
  \end{subfigure}%
  \begin{subfigure}[b]{.2\textwidth}
    \begin{tikzpicture}[comp25,rotated]
      \drawRotatedSpider{}
      \fill [source]
      (D1-5) --                   %
      (D2-2) --                   %
      (D3-10) --                  %
      (D4-8) -- cycle;            %
    \end{tikzpicture}
    \caption{CF-Locking\\ \cite{Bletsch2011a}%
     }\label{fig:spider:cf-locking}
  \end{subfigure}%
  \begin{subfigure}[b]{.2\textwidth}
    \begin{tikzpicture}[comp25,rotated]
      \drawRotatedSpider{}
      \fill [binary]
      (D1-5) --                   %
      (D2-2) --                   %
      (D3-10) --                  %
      (D4-1) -- cycle;            %
    \end{tikzpicture}
    \caption{MoCFI\\ \cite{davi.etal+12}%
     }\label{fig:spider:mocfi}
  \end{subfigure}%
  \begin{subfigure}[b]{.2\textwidth}
    \begin{tikzpicture}[comp25,rotated]
      \drawRotatedSpider{}
      \fill [source]
      (D1-5) --                   %
      (D2-2) --                   %
      (D3-1) --                  %
      (D4-8) -- cycle;            %
    \end{tikzpicture}
    \caption{MIP\\ \cite{Niu2013a}%
     }\label{fig:spider:mpi}
  \end{subfigure}%
  \\
  \begin{subfigure}[b]{.2\textwidth}
    \begin{tikzpicture}[comp25,rotated]
      \drawRotatedSpider{}
      \fill [binary]
      (D1-5) --                   %
      (D2-2) --                   %
      (D3-1) --                  %
      (D4-8) -- cycle;            %
    \end{tikzpicture}
    \caption{kBouncer\\ \cite{kBouncer}%
     }\label{fig:spider:kbouncer}
  \end{subfigure}%
  \begin{subfigure}[b]{.2\textwidth}
    \begin{tikzpicture}[comp25,rotated]
      \drawRotatedSpider{}
      \fill [source]
      (D1-5) --                   %
      (D2-4) --                   %
      (D3-3) --                  %
      (D4-5) -- cycle;            %
    \end{tikzpicture}
    \caption{CF-Restrictor~\cite{pewny.holz+13}%
     }\label{fig:spider:restrictor}
  \end{subfigure}%
  \begin{subfigure}[b]{.2\textwidth}
    \begin{tikzpicture}[comp25,rotated]
      \drawRotatedSpider{}
      \fill [binary]
      (D1-5) --                   %
      (D2-2) --                   %
      (D3-3) --                  %
      (D4-9) -- cycle;            %
    \end{tikzpicture}
    \caption{\CCFIR\\ \cite{CCFIR}%
     }\label{fig:spider:ccfir}
  \end{subfigure}%
  \begin{subfigure}[b]{.2\textwidth}
    \begin{tikzpicture}[comp25,rotated]
      \drawRotatedSpider{}
      \fill [binary]
      (D1-5) --                   %
      (D2-2) --                   %
      (D3-3) --                  %
      (D4-7) -- cycle;            %
    \end{tikzpicture}
    \caption{\binCFI \cite{cfi-cots}%
     }\label{fig:spider:bincfi}
  \end{subfigure}%
  \begin{subfigure}[b]{.2\textwidth}
    \begin{tikzpicture}[comp25,rotated]
      \drawRotatedSpider{}
      \fill [runtime]
      (D1-8) --                   %
      (D2-2) --                   %
      (D3-1) --                  %
      (D4-9) -- cycle;            %
    \end{tikzpicture}
    \caption{ROPecker\\ \cite{ropecker}%
     }\label{fig:spider:ropecker}
  \end{subfigure}%
  \\
  \begin{subfigure}[b]{.2\textwidth}
    \begin{tikzpicture}[comp25,rotated]
      \drawRotatedSpider{}
      \fill [source]
      (D1-5) --                   %
      (D2-2) --                   %
      (D3-3) --                  %
      (D4-7) -- cycle;            %
    \end{tikzpicture}
    \caption{KCoFI\\ \cite{criswell.etal+14}%
     }\label{fig:spider:kcofi}
  \end{subfigure}%
  \begin{subfigure}[b]{.2\textwidth}
    \begin{tikzpicture}[comp25]
      \drawFwdOnlySpider{}
      \fill [source]
      (D1-6) --                   %
      (D2-4) --                   %
      (D3-9) -- cycle;            %
    \end{tikzpicture}
    \caption{SafeDispatch\\ \cite{jang.etal+14}%
     }\label{fig:spider:safedispatch}
  \end{subfigure}%
  \begin{subfigure}[b]{.2\textwidth}
    \begin{tikzpicture}[comp25]
      \drawFwdOnlySpider{}
      \fill [binary]
      (D1-6) --                   %
      (D2-2) --                   %
      (D3-9) -- cycle;            %
    \end{tikzpicture}
    \caption{T-VIP \cite{t-vip}%
     }\label{fig:spider:tvip}
  \end{subfigure}%
  \begin{subfigure}[b]{.2\textwidth}
    \begin{tikzpicture}[comp25,rotated]
      \drawRotatedSpider{}
      \fill [hardware]
      (D1-6) --                   %
      (D2-4) --                   %
      (D3-10) --                  %
      (D4-9) -- cycle;            %
    \end{tikzpicture}
    \caption{HW-asst. CFI\\ \cite{davi.etal+14-hw-cfi}%
     }\label{fig:spider:hardware-cfi}
  \end{subfigure}%
  \begin{subfigure}[b]{.2\textwidth}
    \begin{tikzpicture}[comp25,rotated]
      \drawRotatedSpider{}
      \fill [source]
      (D1-8) --                   %
      (D2-6) --                   %
      (D3-3) --                  %
      (D4-5) -- cycle;            %
    \end{tikzpicture}
    \caption{RockJIT\\ \cite{niu.tan+14-rockjit}%
     }\label{fig:spider:rockjit}
  \end{subfigure}%
  \\
  \begin{subfigure}[b]{.2\textwidth}
    \begin{tikzpicture}[comp25]
      \drawFwdOnlySpider{}
      \fill [source]
      (D1-7) --                   %
      (D2-4) --                   %
      (D3-8) -- cycle;            %
    \end{tikzpicture}
    \caption{\VTV \\ \cite{tice.etal+14}%
     }\label{fig:spider:fcfi-vtv}
  \end{subfigure}%
  \begin{subfigure}[b]{.2\textwidth}
    \begin{tikzpicture}[comp25,rotated]
      \drawRotatedSpider{}
      \fill [binary]
      (D1-6) --                   %
      (D2-2) --                   %
      (D3-3) --                  %
      (D4-8) -- cycle;            %
    \end{tikzpicture}
    \caption{O-CFI\\ \cite{mohan.etal+15}%
     }\label{fig:spider:ocfi}
  \end{subfigure}%
  \begin{subfigure}[b]{.2\textwidth}
    \begin{tikzpicture}[comp25,rotated]
      \drawRotatedSpider{}
      \fill [source]
      (D1-8) --                   %
      (D2-8) --                   %
      (D3-6) --                  %
      (D4-1) -- cycle;            %
    \end{tikzpicture}
    \caption{\CryptoCFI~\cite{mashtizadeh.etal+15}%
     }\label{fig:spider:ccfi}
  \end{subfigure}%
  \begin{subfigure}[b]{.2\textwidth}
    \begin{tikzpicture}[comp25]
      \drawFwdOnlySpider{}
      \fill [binary]
      (D1-6) --                   %
      (D2-6) --                   %
      (D3-3) -- cycle;            %
    \end{tikzpicture}
    \caption{vfGuard \\ \cite{vfguard}%
     }\label{fig:spider:vfguard}
  \end{subfigure}%
  \begin{subfigure}[b]{.2\textwidth}
    \begin{tikzpicture}[comp25]
      \drawFwdOnlySpider{}
      \fill [binary]
      (D1-6) --                   %
      (D2-2) --                   %
      (D3-9) -- cycle;            %
    \end{tikzpicture}
    \caption{VTint\\ \cite{vtint}%
     }\label{fig:spider:vtint}
  \end{subfigure}%
\\
  \begin{subfigure}[b]{.2\textwidth}
    \begin{tikzpicture}[comp25,rotated]
      \drawRotatedSpider{}
      \fill [binary]
      (D1-6) --                   %
      (D2-9) --                   %
      (D3-5) --                   %
      (D4-9) -- cycle;            %
    \end{tikzpicture}
    \caption{PathArmor~\cite{veen.etal+15}%
     }\label{fig:spider:patharmor}
  \end{subfigure}%
  \begin{subfigure}[b]{.2\textwidth}
    \begin{tikzpicture}[comp25,rotated]
      \drawRotatedSpider{}
      \fill [runtime]
      (D1-6) --                   %
      (D2-2) --                   %
      (D3-2) --                   %
      (D4-8) -- cycle;            %
    \end{tikzpicture}
    \caption{CFIGuard\\ \cite{yuan.etal+15}%
     }\label{fig:spider:cfiguard}
  \end{subfigure}%
  \begin{subfigure}[b]{.2\textwidth}
    \begin{tikzpicture}[comp25]
      \drawFwdOnlySpider{}
      \fill [source]
      (D1-6) --                   %
      (D2-2) --                   %
      (D3-9) -- cycle;            %
    \end{tikzpicture}
    \caption{\CFGuard \\ \cite{CFGUARD}%
     }\label{fig:spider:cfguard}
  \end{subfigure}%
  \begin{subfigure}[b]{.2\textwidth}
    \begin{tikzpicture}[comp25]
      \drawFwdOnlySpider{}
      \fill [source]
      (D1-6) --                   %
      (D2-4) --                   %
      (D3-10) -- cycle;            %
    \end{tikzpicture}
    \caption{\VTI \\ \cite{bounov.etal+15}%
     }\label{fig:spider:vti}
  \end{subfigure}%
  \begin{subfigure}[b]{.2\textwidth}
    \begin{tikzpicture}[comp25,rotated]
      \drawRotatedSpider{}
      \fill [source]
      (D1-8) --                 %
      (D2-10) --                 %
      (D3-3) --                 %
      (D4-9) -- cycle;          %
    \end{tikzpicture}
    \caption{Kernel CFI \\ \cite{ge16eurosp}%
     }\label{fig:qspider:ge16}
  \end{subfigure}%
  \caption{CFI implementation comparison: supported control-flows (CF), reported
      performance (RP), static analysis precision: forward (SAP.F) and backward
          (SAP.B). Backward (SAP.B) is omitted for mechanisms that do not
          support back edges.  Color coding of CFI implementations: binary are
          blue, source-based are green, others red.}
 \label{fig:spiderweb}
\end{figure}
}

\subsection{Qualitative Security Guarantees}\label{ss:eval-qualitative} 
\begin{figure}
\captionsetup[subfigure]{justification=centering}
 \centering
  \begin{subfigure}[b]{.2\textwidth}
    \begin{tikzpicture}[comp4]
      \drawQuantSpider{}
      \fill [source]
      (D1-8) --                 %
      (D2-8) --                 %
      (D3-3) --                %
      (D4-9) --                 %
      (D5-8) -- cycle;          %
    \end{tikzpicture}
    \caption{\MCFI \\ \cite{niu.tan+14}%
     }\label{fig:qspider:mcfi}
  \end{subfigure}%
  \begin{subfigure}[b]{.2\textwidth}
    \begin{tikzpicture}[comp4]
      \drawQuantSpider{}
      \fill [source]
      (D1-8) --                 %
      (D2-10) --                 %
      (D3-6) --                %
      (D4-9) --                 %
      (D5-9) -- cycle;          %
    \end{tikzpicture}
    \caption{\piCFI \\ \cite{niu.tan+15}%
     }\label{fig:qspider:picfi}
  \end{subfigure}%
  \begin{subfigure}[b]{.2\textwidth}
    \begin{tikzpicture}[comp3,rotated]
      \drawQRotatedSpider{}
      \fill [source]
      (D1-5) --                 %
      (D2-10) --                 %
      (D3-9) --                 %
      (D4-5) -- cycle;          %
    \end{tikzpicture}
    \caption{\IFCC \\ \cite{tice.etal+14}%
     }\label{fig:qspider:ifcc}
  \end{subfigure}%
  \begin{subfigure}[b]{.2\textwidth}
    \begin{tikzpicture}[comp3]
      \drawQFwdOnlySpider{}
      \fill [source]
      (D1-5) --                 %
      (D2-10) --                 %
      (D3-5) -- cycle;          %
    \end{tikzpicture}
    \caption{\LLVMCFI-3.7\\(2015)}\label{fig:qspider:fecvc}
  \end{subfigure}%
  \begin{subfigure}[b]{.2\textwidth}
    \begin{tikzpicture}[comp4]
      \drawQuantSpider{}
      \fill [binary]
      (D1-9) --                 %
      (D2-4) --                 %
      (D3-10) --                %
      (D4-3) --                 %
      (D5-5) -- cycle;          %
    \end{tikzpicture}
    \caption{\LD \\ \cite{lockdown}%
     }\label{fig:qspider:lockdown}
  \end{subfigure}%
  \caption{Quantitative comparison: control-flows (CF), quantitative security
      (Q), reported performance (RP), static analysis precision: forward (SAP.F) and backward (SAP.B).}
  \label{fig:quantitative-comparison}
\end{figure}
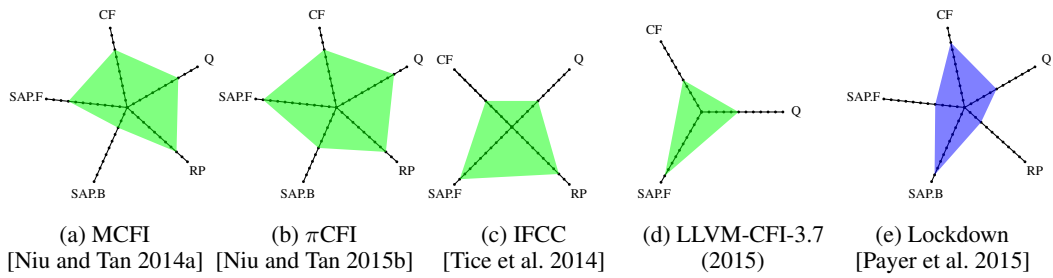

Our
qualitative analysis of prior work and proposed CFI implementations relies on
the classifications of the previous section (cf.~\autoref{s:foundations}) to
provide a higher resolution view of precision and security.
\autoref{fig:spiderweb} summarizes our findings among four dimensions based on
the author's reported results and analysis techniques.
\autoref{fig:quantitative-comparison} presents our verified results for open
source LLVM-based implementations that we have selected.  Further, it adds a
quantitative argument based on our work in \autoref{ss:eval-quantitative}.

In \autoref{fig:spiderweb} the axes and values were calculated as follows.  Note
that (i) the scale of each axis varies based on the number of data points required
and (ii) weaker/slower always scores lower and stronger/faster higher.
Therefore, the area of the spider plot roughly estimates the
security/precision of a given mechanism:
\begin{itemize}
\itemsep-0pt

\item CF: supported control-flow transfers, assigned based on our classification
scheme in \autoref{ss:cflow-classification};

\item RP: reported performance numbers. Performance is quantified on a scale of 1-10 by taking the arctangent of reported runtime overhead and normalizing for high granularity near the median overhead. An implementation with no overhead receives a full score of 10, and one with about 35\% or greater overhead receives a minimum score of 1.

\item SAP.F: static-analysis precision of forward control-flows, assigned based
on our classification in \autoref{sss:sap}; and

\item SAP.B: static-analysis precision of backward control-flows, assigned based
on our classification in \autoref{sss:sap}.

\end{itemize}

The shown CFI implementations are ordered chronologically by publication year,
and the colors indicate whether a CFI implementation works on the binary-level
(blue), relies on source-code (green), or uses other mechanisms (red), such as
hardware implementations.

Our classification and categorization efforts for reported performance
were hindered by methodological variances in benchmarking.  Experiments were
conducted on different machines, different operating systems, and also different
or incomplete benchmark suites.  Classifying and categorizing static analysis
precision was impeded by the
high level, imprecise descriptions of the implemented static analysis by various
authors.  Both of these
impediments, naturally, are sources of imprecision in our evaluation.
Comprehensive protection through CFI requires the validation of both forward and
backward branches.  This requirement means that the reported performance impact
for forward-only approaches (i.e., SafeDispatch, T-VIP, VTV, \IFCC, vfGuard, and
VTint) is restricted to partial protection.  The performance impact for backward
control-flows must be considered as well, when comparing these mechanisms to
others with full protection.

CFI mechanisms satisfying SAP.B.2, i.e., using a shadow stack to obtain
high precision for backward control-flows are: original CFI~\cite{abadi.etal+05},
MoCFI~\cite{davi.etal+12}, HAFIX~\cite{davi.etal+14-hw-cfi,arias.etal+15-hafix},
and \LD~\cite{lockdown}.
\PathArmor emulates a shadow stack through validating the last-branch register
(LBR).

Increasing the precision of static analysis techniques that validate whether
any given
control-flow transfer corresponds to an edge in the CFG decreases the
performance of the CFI mechanism.  Most implementations choose to combine
precise results of static
analysis into an equivalence class.  Each such equivalence class receives a
unique identifier, often referred to as a label, which the CFI enforcement
component validates at runtime.  By not using a shadow stack, or any other
comparable high-precision backward control-flow transfer validation mechanism,
even high precision forward control-flow transfer static analysis becomes
imprecise due to labeling.  The explanation for this loss in precision is
straightforward: to validate a control-flow transfer, all callers of a function
need to carry the same label.  Labeling, consequently, is a substantial source
of imprecision (see~\autoref{ss:eval-quantitative} for more details).  The
notable exception in this case is \piCFI, which uses dynamic information, to
activate pre-determined edges, dynamically enabling
high-resolution,
precise control-flow graph (somewhat analogous to dynamic points-to
sets~\cite{Mock2001}.  Borrowing a term from information-flow
control~\cite{Sabelfeld2003}, \piCFI can, however, suffer from \emph{label
creep} by accumulating too many labels from the static CFG.

CFI implementations introducing imprecision via labeling are: the original CFI
paper~\cite{abadi.etal+05}, control-flow locking~\cite{Bletsch2011a},
CF-restrictor~\cite{pewny.holz+13}, \CCFIR~\cite{CCFIR},
\MCFI~\cite{niu.tan+14}, KCoFI~\cite{Criswell2014}, and
RockJIT~\cite{niu.tan+14-rockjit}.

According to the criteria established in analyzing points-to precision, we find
that at the time of this writing, \piCFI~\cite{niu.tan+15} offers the highest
precision due to leveraging dynamic points-to information.  \piCFI's
predecessors, RockJIT~\cite{niu.tan+14-rockjit} and \MCFI~\cite{niu.tan+14},
already offered a high precision due to the use of context-sensitivity in
the form of types.  Ideal \PathArmor also scores well when subject to our
evaluation: high-precision in both directions, forward and backward, but
is hampered by limited hardware resources (LBR size) and restricting protection to
the main executable (i.e., trusting libraries).
\LD~\cite{lockdown}
offers high precision on the backward edges but derives its equivalence
classes from the number of libraries used in an application and is therefore
inherently limited in the precision of the forward edges.
\IFCC~\cite{tice.etal+14} offers variable static analysis granularity.  On
the one hand, \IFCC describes a Full mode that uses type information,
similar to \piCFI and its predecessors. On the other hand, \IFCC mentions
less precise modes, such as using a single set for all destinations, and
separating by function arity.  With the exception of
Hypersafe~\cite{wang10oakland}, all other evaluated CFI implementations with
supporting academic publications offer lower precision of varying degrees,
at most as precise as SAP.F.3.

\subsection{Quantitative Security Guarantees}\label{ss:eval-quantitative}

Quantitatively assessing how much security a CFI mechanism provides is
challenging as attacks are often program dependent and different
implementations might allow different attacks to succeed.  So far, the only
existing quantitative measure of the security of a CFI implementation is Average
Indirect Target Reduction (AIR).  Unfortunately, AIR is known to be a
weak proxy for security~\cite{tice.etal+14}.  A more meaningful metric must
focus on the number of targets (i.e., number of equivalence classes) available
to an attacker.  Furthermore, it should recognize that smaller classes are more
secure, because they provide less attack surface.  Thus, an implementation with a
small number of large equivalence classes is more vulnerable than an
implementation with a large number of small equivalence classes.

One possible metric is the product of the number of equivalence classes (EC) and
the inverse of the size of the largest class (LC), see
\autoref{eq:quant-metric}. Larger products indicate a more secure
mechanism as the product increases with the number of equivalence classes and
decreases with the size of the largest class.  More equivalence classes means
that each class is smaller, and thus provides less attack surface to an
adversary.  Controlling for the size of the largest class attempts to control
for outliers, e.g., one very large and thus vulnerable class and many smaller
ones.  A more sophisticated version would also consider the usability and
functionality of the sets. Usability considers whether or not they are located
on an attacker accessible ``hot'' path, and if so how many times they are used.
Functionality evaluates the quality of the sets, whether or not they include
``dangerous'' functions like mprotect. A large equivalence class that is pointed
to by many indirect calls on the hot path poses a higher risk because it is more
accessible to the attacker.

\begin{equation}\label{eq:quant-metric}
{EC}*\frac{1}{LC} = Quantitative Security
\end{equation}

This metric is not perfect, but it allows a meaningful direct comparison of the
security and precision of different CFI mechanisms, which AIR does not.  The
gold standard would be adversarial analysis.  However, this currently requires a
human to perform the analysis on a per-program basis.  This leads to a large
number of methodological issues: how many analysts, which programs and inputs,
how to combine the results, etc.  Such a study is beyond the scope of this work,
which instead uses our proposed metric which can be measured programatically.

This section measures the number and sizes of sets to allow a meaningful, direct
comparison of the security provided by different implementations.  Moreover, we
report the dynamically observed number of sets and their sizes.  This quantifies
the maximum achievable precision from the implementations' CFG analysis, and
shows how over-approximate they were for a given execution of the program.

\subsubsection{Implementations}\label{sss:quant-implementations}

We evaluate four compiler-based, open-source CFI mechanisms \IFCC, \LLVMCFI,
\MCFI, and \piCFI.  
For \IFCC and \MCFI we also evaluated the different
analysis techniques available in the implementation.  Note that we evaluate
two different versions of \LLVMCFI, the first release in LLVM 3.7 and the
second, highly modified version in LLVM 3.9.  In addition to the compiler-based
solutions, we also evaluate \LD, which is a binary-based CFI implementation.

\MCFI and \piCFI already have a built-in reporting mechanism. For the other
mechanisms we extend the instrumentation pass and report the number and size of
the produced target sets.  We then used the implementations to compile, and for
\piCFI run, the \SpecCPU benchmarks to produce the data we report
here.
\piCFI must be run because it does dynamic target activation.  This does tie our
results to the ref data set for \SpecCPU, because as with any dynamic
analysis the results will depend on the input.

\IFCC\footnote{Note that the \IFCC patch was pulled by the authors and will be
replaced by \LLVMCFI.} comes with four different CFG analysis techniques:
\emph{single}, \emph{arity}, \emph{simplified}, and \emph{full}.  \emph{Single}
creates only one equivalence class for the entire program, resulting in the weakest
possible CFI policy. %
\emph{Arity} groups functions into equivalence classes based on their number of
arguments.  \emph{Simplified} improves on this by recognizing three types of
arguments: composite, integer, or function pointer. \emph{Full} considers the
precise return type and types of each argument.  We expect full to yield the
largest number of equivalence classes with the smallest sizes, as it performs
the most exact distribution of targets.

Both \MCFI and \piCFI rely on the same underlying static analysis. %
The authors claim that disabling tail calls is
the single most important precision enhancement for their CFG
analysis~\cite{the-github-readme-for-mcfi}. We measure the impact of this option
on our metric. \MCFI and \piCFI are also unique in that their policy and
enforcement mechanisms consider backward edges as well as forward edges.  When
comparing to other implementations, we only consider forward edges.  This ensures
direct comparability for the number and size of sets.  The results
for backward edges are presented as separate entries in the figures.

As of LLVM 3.7, \LLVMCFI could not be directly compared to the other CFI
implementations because its policy was strictly more limited.  Instead of
considering all forward, or all forward and backward edges, \LLVMCFI 3.7 focused
on virtual calls and ensures  that virtual, and non-virtual calls are performed
on objects of the correct dynamic type.  As of LLVM 3.9, \LLVMCFI has added
support for all indirect calls. Despite these differences, we show the full
results for both \LLVMCFI implementations in all tables and graphs.

\LD is a CFI implementation that operates on compiled binaries and supports
the instrumentation of dynamically loaded code. To protect backward edges,
\LD enforces a shadow stack. For the forward edge, it instruments libraries at
runtime, creating one equivalence class per library.  Consequently, the set size
numbers are of the greatest interest for \LD.  \LD's precision depends on symbol
information, allowing indirect calls anywhere in a particular library if it is
stripped.  Therefore, we only report the set sizes for non-stripped libraries
where \LD is more precise.

To collect the data for our lower bound, we wrote an LLVM pass.  This pass
instruments the program to collect and report the source line for each indirect
call, the number of
different targets for each indirect call, and the number of times each of
those targets was used.  This data is collected at runtime.  Consequently,
it represents only a subset of all possible indirect calls and targets that are
required for the sample input to run.  As
such, we use it to present a lower bound on the number of equivalence sets
(i.e. unique indirect call sites) and size of those sets (i.e. the number of
different locations called by that site).

\subsubsection{Results}

We conducted three different quantitative evaluations in line with our proposed
metric for evaluating the overall security of a CFI mechanism and our lower
bound. For \IFCC, \LLVMCFI (3.7 and 3.9), and \MCFI it is sufficient to compile
the \SpecCPU benchmarks as they do not dynamically change their equivalence
classes.  \piCFI uses dynamic information, so we had to run the \SpecCPU
benchmarks.  Similarly, Lockdown is a binary CFI implementation that only
operates at run time.  We highlight the most interesting results in
\autoref{fig:quantitative-comparison}, see \autoref{tbl:quant-full-results} in
\autoref{sec:appendix} for the full data set.

\begin{figure*}
  \centering
  \includegraphics[width=\textwidth]{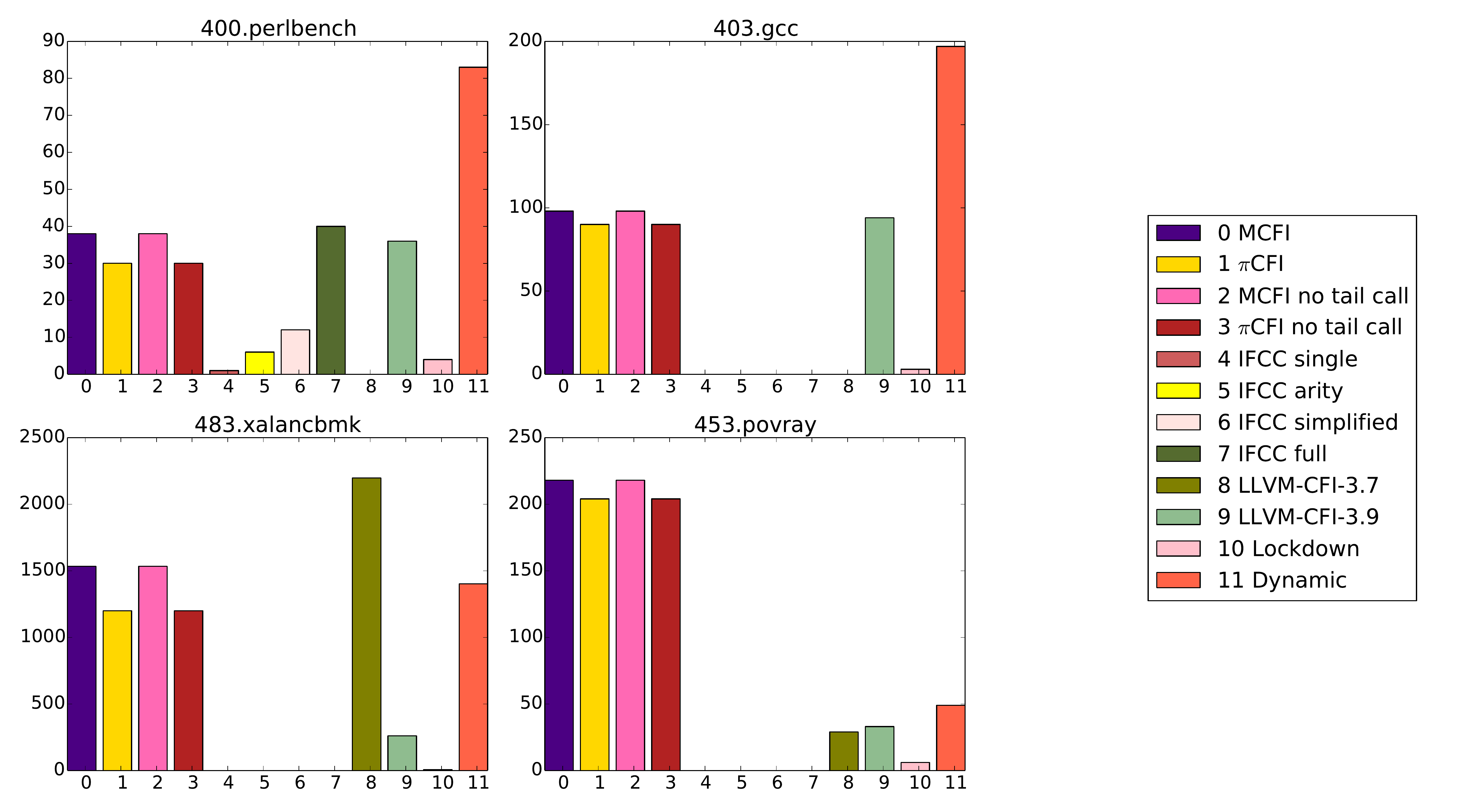}
  \caption{Total number of forward-edge equivalence classes when running SPEC CPU2006 (higher is better).}
  \label{fig:equivalence-classes}
\end{figure*}

\autoref{fig:equivalence-classes} shows the number of equivalence classes for
the five CFI implementations that we evaluated, as well as their
sub-configurations. As advertised, \IFCC \emph{Single} only creates one
equivalence class.
This \IFCC mode offers the least precision of any implementation measured.
The other \IFCC analysis modes only had a
noticeable impact for perlbench and soplex.  Indeed, on the sjeng benchmark all
four analysis modes produced only one equivalence class.

On forward edges, \MCFI and \piCFI are more precise than \IFCC in all cases
except for perlbench where they are equivalent.  \LLVMCFI 3.9 is more precise
than \IFCC while being less precise than \MCFI.  \MCFI and \piCFI are the only
implementations to consider backward edges, so no comparison
with other mechanisms is possible on backward edge precision.  Relative to each
other, \piCFI's dynamic information decreases the number of equivalence classes
available to the attacker by 21.6\%.  The authors of \MCFI and \piCFI recommend
disabling tail calls to improve CFG precision. This only impacts the number of
sets that they create for backward edges, not forward edges,
see~\autoref{sec:appendix}.  As such this compiler flag does not impact
most CFI implementations, which rely on a shadow stack for backward edge
security.

\begin{figure*}
  \centering
  \includegraphics[width=\textwidth]{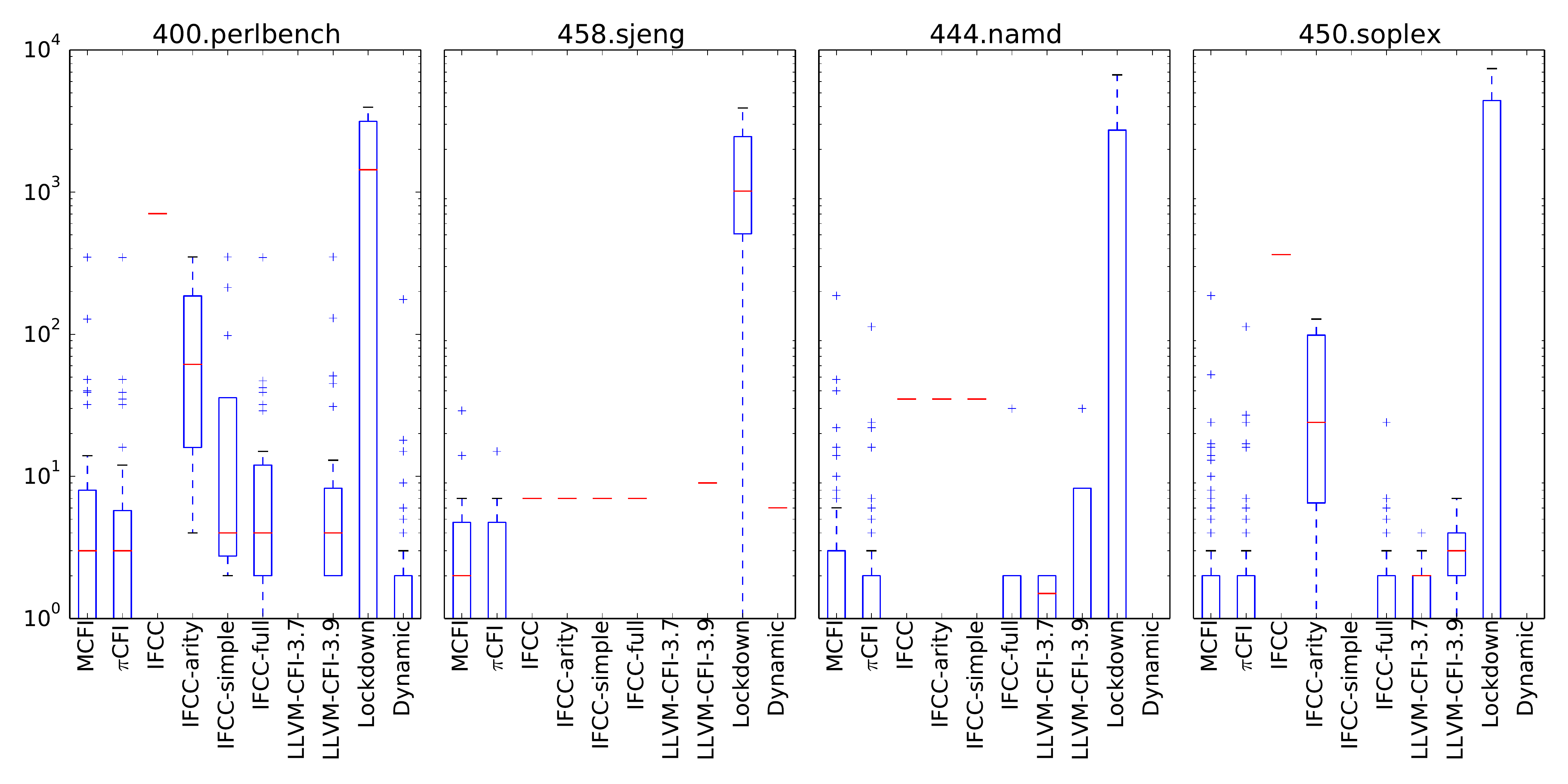}
  \caption{Whisker plot of equivalence class sizes for different mechanisms when running SPEC CPU2006.
            (Smaller is Better)}
  \label{fig:set-sizes}
\end{figure*}

\LLVMCFI 3.7 creates a number of equivalence classes equal to the number of
classes used in the \C++ benchmarks.  Recall that it only provides support for
a subset of indirect control-flow transfer types.  However, we present the results in
\autoref{fig:equivalence-classes} and \autoref{fig:set-sizes} to show the
relative cost of protecting vtables in \C++ relative to protecting all indirect call
sites.

We quantify the set sizes for each of the four implementations
in~\autoref{fig:set-sizes}. We show box and whisker graphs of the set sizes for
each implementation.  The red line is the median set size and a smaller median
set size indicates more secure mechanisms. The blue box extends from the
\nth{25} percentile to the \nth{75}, smaller boxes indicate a tight grouping
around the median.  An implementation might have a low median, but large boxes
indicate that there are still some large equivalence classes for an attacker
to target.  The top whisker extends from the top of the box for 150\% of the
size of the box.  Data points beyond the whiskers are considered outliers and
indicate large sets.  This plot format allows an intuitive understanding of the
security of the distribution of equivalence class sizes.  Lower medians and
smaller boxes are better.  Any data points above the top of the whisker show
very large, outlier equivalence classes that provide a large attack surface for
an adversary.

Note that \IFCC only creates a single equivalence class for xalancbmk and namd
(except for the Full configuration on namd which is more precise). Entries with
just a single equivalence class are reported as only a median.  \IFCC data
points allow us to rank the different analysis methods, based on the results for
benchmarks where they actually impacted set size: perlbench and soplex.  In
increasing order of precision (least precise to most precise) they are:
\emph{single}, \emph{arity}, \emph{simplified}, and \emph{full}. %
This does not
necessarily mean that the more precise analysis methods are more secure,
however.  For perlbench the more precise methods have outliers at the same level
as the median for the least precise (i.e., \emph{single}) analysis.  For soplex
the outliers are not as bad, but the \emph{full} outlier is the same size as the
median for \emph{arity}.  While increasing the precision of the underlying CFG
analysis increases the overall security, edge cases can cause the incremental
gains to be much smaller than anticipated.

The \MCFI forward-edge data points highlight this.  The \MCFI median is always
smaller than the \IFCC median.  However, for all the benchmarks where both ran,
the \MCFI outliers are greater than or equal to the largest \IFCC set.  From a
quantitative perspective, we can only confirm that \MCFI is at least as secure
as \IFCC.  The effect of the outlying large sets on relative security remains
an open question, though it seems likely that they provide opportunities for an
attacker.

\LLVMCFI 3.9 presents an interesting compromise. As the full set of whisker
plots in~\autoref{sec:appendix} shows, it has fewer outliers.  However, it also
has, on average, a greater median set size.  Given the open question of the
importance of the outliers, \LLVMCFI 3.9 could well be more secure in practice.

\LLVMCFI 3.7 - which only protects virtual tables - sets do not have extreme outliers.
Additionally, \autoref{fig:set-sizes} shows that the equivalence classes that
are created have a low variance, as seen by the more compact whisker plots that
lack the large number of outliers present for other techniques.  As such,
\LLVMCFI 3.7 does not suffer from the edge cases that effect more general
analyzes.
\LD consistently has the largest set sizes, as expected because it only
creates one equivalence class per library and the SPEC CPU2006 benchmarks are
optimized to reduce the amount of external library calls. These sets are up to an order
of magnitude larger than compiler techniques.  However, \LD isolates faults into
libraries as each library has its independent set of targets compared to a
single set of targets for other binary-only approaches like \CCFIR and \binCFI.
The lower bound numbers were measured dynamically, and as such encapsulate
a subset of the actual equivalence sets in the static program.  Further, each
such set
is at most the size of the static set.  Our lower bound thus provides a proxy
for an ideal CFI implementation in that it is perfectly precise for each run.
However, all of the \IFCC
variations report fewer equivalence classes than our dynamic bound.

The whisker plots for our dynamic lower bound in \autoref{fig:set-sizes} show
that some of the \SpecCPU benchmarks inherently have outliers in their set
sizes.  For perlbench, gcc, gobmk, h264ref, omnetpp, and xalancbmk our dynamic
lower bound and the static set sizes from the compiler-based implementations all
have a significant number of outliers.  This provides quantitative backing to
the intuition that some code is more amenable to protection by CFI.  Evaluating what coding
styles and practices make code more or less amenable to CFI is out of scope
here, but would make for interesting future work.

Note that for namd and soplex in \autoref{fig:set-sizes} there is no visible
data for our dynamic lower bound because all the sets had a single element.
This means the median size is one which is too low to be visible.  For all other
mechanisms no visible data means the mechanism was incompatible with the
benchmark.

\subsection{Previous Security Evaluations and Attacks}

Evaluating the security of a CFI implementation is challenging because exploits
are program dependent and simple metrics do not cover the security of a
mechanism. The Average Indirect target Reduction (AIR) metric~\cite{cfi-cots}
captures the average reduction of allowed targets, following the idea that an
attack is less likely if fewer targets are available.  This metric and variants
were then used
to measure new CFI implementations, generally reporting high numbers of more
than $99\%$. Such high numbers give the illusion of relatively high security
but, e.g., if a binary has 1.8 MB of executable code (the size of the glibc on
Ubuntu 14.04), then an AIR value of $99.9\%$ still allows 1,841 targets, likely
 enough for an arbitrary attack. A similar alternative metric to evaluate
CFI effectiveness is the gadget reduction metric~\cite{niu.tan+14}.
Unfortunately, these simple relative metrics give, at best, an intuition
for security and we argue that a more rigorous metric is needed.

A first set of attacks against CFI implementations targeted
\emph{coarse-grained} CFI that only had 1-3 equivalence
classes~\cite{out-of-control, ROP-CFI:Davi, ROP-CFI:Wagner}.  These attacks show
that equivalence classes with a large number of targets allow an attacker to
execute code and system calls, especially if return instructions are allowed to
return to any call site.

Counterfeit Object Oriented Programming (COOP)~\cite{schuster.etal+15}
introduced the idea that whole \C++ methods can be used as gadgets to implement
Turing-complete computation. Virtual
calls in \C++ are a specific type of indirect function calls that are dispatched
via vtables, which are arrays of function pointers. COOP shows that an
attacker can construct counterfeit objects and, by reusing existing vtables,
perform arbitrary computations. This attack shows that indirect calls requiring
another level-of-indirection (e.g., through a vtable) must have additional
checks that consider the types at the language level for the security check as
well.

Control Jujutsu~\cite{Evans15Jujutsu} extends the existing attacks to so-called
fine-grained CFI by leveraging the imprecision of points-to analysis. This
work shows that common software engineering practices like modularity (e.g.,
supporting plugins and refactoring) force points-to analysis to merge several
equivalence classes. This imprecision results in target sets that are large
enough for arbitrary computation.

Control-Flow Bending~\cite{cfbending} goes one step further and shows that
attacks against ideal CFI are possible. Ideal CFI assumes that a precise CFG
is available that is not achievable in practice, i.e., if any edge would be
removed then the program would fail. Even in this configuration attacks are
likely possible if no shadow stack is used, and sometimes possible even if a
shadow stack is used.

Several attacks target data structures used by CFI mechanisms.
StackDefiler~\cite{conti.etal+15} leverages the fact that many CFI mechanisms
implement the enforcement as a compiler transformation. Due to this high-level
implementation and the fact that the optimization infrastructure of the compiler
is unaware of the security aspects, an optimization might choose to spill
registers that hold sensitive CFI data to the stack where it can be modified by
an attack~\cite{abadi.etal+05-theory-of-cfi}. Any CFI mechanism will rely on
some runtime data structures that are sometimes writeable (e.g., when \MCFI
loads new libraries and merges existing sets). Missing the
Point~\cite{evans.etal+15} shows that ASLR might not be enough to hide this
secret data from an adversary.

\section{Performance}\label{sec:perf}
While the security properties of CFI (or the lack thereof for some mechanisms)
have received most scrutiny in the academic literature, performance
characteristics play a large part in determining which CFI mechanisms are likely
to see adoption and which are not.  Szekeres et al.~\cite{szekeres.etal+13}
surveyed mitigations against memory corruption and found that mitigations with
more than 10\% overhead do not tend to see widespread adoption in production
environments and that overheads below 5\% are desired by industry practitioners.

Comparing the performance characteristics of CFI mechanisms is a non-trivial
undertaking. Differences in the underlying hardware, operating system, as well as
implementation and benchmarking choices prevents apples-to-apples comparison
between the performance overheads reported in the literature.  For this reason,
we take a two-pronged approach in our performance survey: for a number of
publicly available CFI mechanisms, we measure performance directly on the same
hardware platform and, whenever possible, on the same operating system, and
benchmark suite.  Additionally, we tabulate and compare the performance results
reported in the literature.

We focus on the aggregate cost of CFI enforcement. For a detailed survey of
the performance cost of protecting backward edges from callees to
callers we refer to the recent, comprehensive survey
by~\citet{dang-shadow_stacks}.

\subsection{Measured CFI Performance}
\label{sec:measured-perf}

\paragraph{Selection Criteria} It is infeasible to replicate the reported
performance overheads for all major CFI mechanisms. Many implementations are not
publicly available or require substantial modification to run on modern versions
of Linux or Windows. We therefore focus on recent, publicly available,
compiler-based CFI mechanisms.

Several compiler-based CFI mechanisms share a common lineage. \LLVMCFI, for
instance, improves upon \IFCC, \piCFI improves upon
\MCFI, and \VTI is an improved version of \SD. In those cases, we opted to
measure the latest available version and rely on reported performance numbers
for older versions.

\paragraph{Method} Most authors use the \SpecCPU\ benchmarks to report the
overhead of their CFI mechanism. We follow this trend in our own replication
study. All benchmarks were compiled using the {\tt -O2} optimization level.  The
benchmarking system was a Dell PowerEdge T620 dual processor server having 64GiB
of main memory and two Intel Xeon E5-2660 CPUs running at 2.20 GHz.
To reduce benchmarking noise, we ran the tests on an otherwise idle system and
disabled all dynamic frequency and voltage scaling features.  Whenever possible,
we benchmark the implementations under 64-bit Ubuntu Linux 14.04.2 LTS. The CFI
mechanisms were baselined against  the compiler they were implemented on top of:
\VTV\ on GCC 4.9, \LLVMCFI\ on LLVM 3.7 and 3.9, \VTI\ on LLVM 3.7, \MCFI\ on LLVM 3.5, \piCFI\ on LLVM 3.5.  Since
\CFGuard\ is part of Microsoft Visual \C++ Compiler, MSVC, we used  MSVC 19 to compile and run \SpecCPU\ on a pristine 64-bit Windows 10
installation. We report the geometric mean overhead averaged over three
benchmark runs using the reference inputs in \autoref{tbl:measured-perf}.

Some of the CFI mechanisms we benchmark required link-time optimization, LTO,
which allows the compiler to analyze and optimize across compilation units.
\LLVMCFI and \VTI both require LTO, so for these mechanisms, we report overheads
relative to a baseline \SpecCPU run that also had LTO enabled. The increased
optimization scope enabled by LTO can allow the compiler to perform additional
optimizations such as de-virtualization to lower the cost of CFI enforcement. On
the other hand, LLVM's LTO is less practical than traditional, separate
compilation, e.g., when compiling large, complex code bases.  To measure
the \piCFI mechanism, we applied the author's patches\footnote{The patches are
available at: \url{https://github.com/mcfi/MCFI/tree/master/spec2006}.} for 7 of the SPEC CPU2006
benchmarks to remove coding constructs
that are not handled by \piCFI's control-flow graph analysis~\cite{niu.tan+14}. Likewise, the authors of \VTI provided a patch for the xalancbmk benchmark. It updates code that casts an object instance to its
sibling class, which can cause a CFI violation. We found these patches for hmmer, povray, and xalancbmk to also be necessary for \LLVMCFI 3.9, which otherwise reports a CFI violation on these benchmarks. \VTI was run
in interleaved vtable mode which provides the best performance according to its
authors~\cite{bounov.etal+15}.

\setlength{\tabcolsep}{2.6pt} %
\renewcommand{\arraystretch}{1} %

\begin{sidewaystable}[th]
  \small
	\centering
	\caption{Measured and reported CFI performance overhead (\%) on the \SpecCPU
      benchmarks. The programming language of each benchmark is indicated in
parenthesis: C(C), C++(+), Fortran(F). CF in a cell indicates we were unable to
build and run the benchmark with CFI enabled. Blank cells mean that no results
were reported by the original authors or that we did not attempt to run the
benchmark. Cells with bold fonts indicate 10\% or more overhead, ntc stands for
no tail calls.}
	\label{tbl:measured-perf}
	\begin{tabular}{@{}l|rrrrrrr|rrrrrrrrrr@{}}
		\toprule
      Benchmark & \multicolumn{7}{c|}{Measured Performance} & \multicolumn{10}{c}{Reported Performance} \\

                & \VTV & \multicolumn{2}{c}{\LLVMCFI} & \VTI & \CFGuard & \piCFI & \piCFI  & \VTV & \VTI & \piCFI & \IFCC & \MCFI & \PathArmor & \LD & \CryptoCFI & \ROPecker & \binCFI \\
    Version  &      & \multicolumn{1}{c}{3.7} & \multicolumn{1}{c}{3.9}     &  &          &        &  &      &    &      &   &       &            &     &            &           &         \\
       Options  &      & \multicolumn{1}{c}{LTO} &\multicolumn{1}{c}{LTO} &  \multicolumn{1}{c}{LTO}  &          &        &  \multicolumn{1}{c|}{ntc} &      &   \multicolumn{1}{c}{LTO}   &      &  \multicolumn{1}{c}{LTO}  &       &            &     &            &           &         \\
      \midrule
		\gray
    400.perlbench(C)  &            &  & 2.4 &    &               & 8.2           & 5.3           &               &      & 5.0           & 1.9  & 5.0           & \textbf{15.0} & \textbf{150.0} &                & 5.0           & \textbf{12.0} \\
    401.bzip2(C)      &           &   & -0.7 &  & -0.3          & 1.2           & 0.8           &               &      & 1.0           &      & 1.0           & 0.0           & 8.0            & 5.0            & 0.0           & -9.0          \\
		\gray
    403.gcc(C)        &           &   & CF &   &               & 6.1           & \textbf{10.5} &               &      & 4.5           &      & 4.5           & 9.0           & \textbf{50.0}  &                & 3.0           & 4.5           \\
    429.mcf(C)        &           &   & 3.6 &   & 0.5           & 4.0           & 1.8           &               &      & 4.0           &      & 4.0           & 1.0           & 2.0            & \textbf{10.0}  & 1.0           & 0.0           \\
		\gray
    445.gobmk(C)      &           &   & 0.2 &   & -0.2          & \textbf{11.4} & \textbf{11.8} &               &      & 7.5           &      & 7.0           & 0.0           & \textbf{43.0}  & \textbf{50.0}  & 1.0           & \textbf{15.0} \\
    456.hmmer(C)      &            &   & 0.1 &   & 0.7           & 0.1           & -0.1          &               &      & 0.0           &      & 0.0           & 1.0           & 3.0            & \textbf{10.0}  & 0.0           & -0.5          \\
		\gray
    458.sjeng(C)      &            &   & 1.6 &  & 3.4           & 8.4           & \textbf{11.9} &               &      & 5.0           &      & 5.0           & 0.0           & \textbf{80.0}  & \textbf{40.0}  & 0.0           & -2.5          \\
    464.h264ref(C)    &           &   & 5.3 &   & 5.4           & 7.9           & 8.3           &               &      & 6.0           &      & 6.0           & 1.0           & \textbf{43.0}  & \textbf{45.0}  & 1.0           & \textbf{28.0} \\
		\gray
    462.libquantum(C) &            &   & -6.9 &   &               & -3.0          & -1.0          &               &      & -0.3          &      & 0.0           & 3.0           & 5.0            & \textbf{10.0}  & 0.0           & -0.5          \\
    471.omnetpp(+)    & 5.8           & -1.9 & CF & CF   & 3.8           & 6.7           & \textbf{18.8} & 8.0           & 1.2  & 5.0           & -1.2 & 5.0           &               &                &                & 2.0           & \textbf{45.0} \\
		\gray
    473.astar(+)      & 3.6           & -0.3 & 0.9 & 1.6  & 0.1           & 2.0           & 2.9           & 2.4           & 0.1  & 4.0           & -0.2 & 3.5           &               & \textbf{17.0}  & \textbf{75.0}  & 0.0           & \textbf{14.0} \\
    483.xalancbmk(+)  & \textbf{24.0} & 7.1 & 7.2  & 3.7  & 5.5           & \textbf{10.3} & \textbf{17.6} & \textbf{19.2} & 1.4  & 7.0           & 3.1  & 7.0           &               & \textbf{118.0} & \textbf{170.0} & \textbf{15.0} &               \\
		\gray
      \midrule
		410.bwaves(F)     &               &   &  &      &               &               &               &               &      &               &      &               &               & 1.0            &                &               &               \\
		416.gamess(F)     &               &   &  &      &               &               &               &               &      &               &      &               &               & \textbf{11.0}  &                &               &               \\
		\gray
    433.milc(C)       &            &   & 0.2  &  & 2.0           & -1.7          & 1.4           &               &      & 2.0           &      & 2.0           & 4.0           & 8.0            &                &               & 2.5           \\
    434.zeusmp(F)     &               &   &  &      &               &               &               &               &      &               &      &               &               & 0.0            &                &               &               \\
		\gray
		435.gromacs(C,F)    &               &  &   &      &               &               &               &               &      &               &      &               &               & 1.0            &                &               &               \\
		436.cactusADM(C,F)  &               &  &   &      &               &               &               &               &      &               &      &               &               & 0.0            &                &               &               \\
		\gray
		437.leslie3d(F)   &               &   &  &      &               &               &               &               &      &               &      &               &               & 1.0            &                &               &               \\
    444.namd(+)       & -0.1          & -0.2 & 0.1 & -0.3 & 0.1           & -0.3          & -0.5          &               &      & -0.5          & -0.2 & -0.5          &               & 3.0            &                &               & -2.0          \\
		\gray
    447.dealII(+)     & 0.7           & CF & 7.9 & CF   & -0.1          & 5.3           & 4.4           &               &      & 4.5           & -2.2 & 4.5           &               &                &                &               &               \\
    450.soplex(+)     & 0.5           & 0.5 & -0.3 & -0.6 & 2.3           & -0.7          & 0.9           &               & -0.7 & -4.0          & -1.7 & -4.0          &               & \textbf{12.0}  &                &               & 3.5           \\
		\gray
    453.povray(+)     & -0.6          & 1.5 & 8.9 & 2.0  & \textbf{10.8} & \textbf{11.3} & \textbf{17.4} &               &      & \textbf{10.5} & 0.2  & \textbf{10.0} &               & \textbf{90.0}  &                &               & \textbf{37.0} \\
		454.calculix(C,F)   &               &   &  &      &               &               &               &               &      &               &      &               &               & 3.0            &                &               &               \\
		\gray
		459.gemsFDTD(F)   &               &   &  &      &               &               &               &               &      &               &      &               &               & 7.0            &                &               &               \\
		465.tonto(F)      &               &   &  &      &               &               &               &               &      &               &      &               &               & \textbf{19.0}  &                &               &               \\
		\gray
		470.lbm(C)        &            &   & -0.2 & & 4.2           & -0.2          & -0.5          &               &      & 1.0           &      & 1.0           & 0.0           & 2.0            &                &               & -2.5          \\
		482.sphinx3(C)    &            &  & -0.8 & & -0.1          & 0.7           & 2.4           &               &      & 1.5           &      & 1.5           & 3.0           & 8.0            &                &               & 0.5           \\
      \midrule
      Geo Mean   & 4.6           & 1.1  & 4.4 & 1.3  & 2.3           & 4.0           & 5.8           & 9.6           & 0.5  & 3.2           & -0.3 & 2.9           & 3.0           & \textbf{20.0}  & \textbf{45.0}  & 2.6           & 8.5           \\
      \bottomrule
	\end{tabular}
\end{sidewaystable}

\afterpage{\clearpage}

\paragraph{Results}
Our performance experiments show that recent, compiler-based CFI mechanisms
have mean overheads in the low single digit range.  Such low overhead is well
within the threshold
for adoption specified by~\cite{szekeres.etal+13} of 5\%.  This dispenses with
the
concern that CFI enforcement is too costly in practice compared to alternative
mitigations including those based on randomization~\cite{larsen.etal+14}.
Indeed, mechanisms such as \CFGuard, \LLVMCFI, and \VTV are implemented in
widely-used compilers, offering some level of CFI enforcement to practitioners.

We expect CFI mechanisms that are limited to virtual method calls---\VTV, \VTI,
\LLVMCFI 3.7--- to have lower mean overheads than those that also protect
indirect function calls such as \IFCC.  The return protection mechanism used by
\MCFI should introduce additional overhead, and \piCFI's runtime policy
ought to result in a
further marginal increase in overhead.  In practice, our results show that
\LLVMCFI 3.7 and \VTI are the fastest, followed by \CFGuard, \piCFI, and \VTV.
The reported numbers for \IFCC when run in \emph{single} mode show that it
achieves -0.3\%, likely due to cache effects. Although our
measured overheads are not directly comparable with those reported by the
authors of the seminal CFI paper, we find that researchers have managed to
improve the precision while lowering the cost\footnote{Non-CFI related hardware
improvements,
such as better branch prediction~\cite{rohou.etal+15-branch-predict}, also
help to reduce performance overhead.} of enforcement as the result of a decade
worth of research into CFI enforcement.

The geometric mean overheads do not tell the whole story, however.  It is
important to look closer at the performance impact on benchmarks that execute a
high number of indirect branches. Protecting the xalancbmk, omnetpp, and povray
\C++ benchmarks with CFI generally incurs substantial overheads.  %
All benchmarked
CFI mechanisms had above-average overheads on xalancbmk. \LLVMCFI and \VTV,
which take virtual call semantics into account, were particularly affected. On
the other hand, xalancbmk highlights the merits of the recent virtual table
interleaving mechanism of \VTI which has a relatively low 3.7\% overhead (vs.
1.4\% reported) on this challenging benchmark.

Although povray is written in \C++, it makes few virtual method calls
\cite{vtint}. However, it performs a large number of indirect calls. %
The CFI mechanisms which
protect indirect calls---\piCFI, and \CFGuard---all incur high performance
overheads on povray.  Sjeng and h264ref also include a high number of indirect
calls which again result in non-negligible overheads particularly when using
\piCFI with tail calls disabled to improve CFG precision.  The hmmer, namd, and
bzip2 benchmarks on the other hand show very little overhead as they do not
execute a high number of forward indirect branches of any kind. Therefore these
benchmarks are of little value when comparing the performance of various CFI
mechanisms.

Overall, our measurements generally match those reported in the literature. The
authors of \VTV~\cite{tice.etal+14} only report overheads for the three \SpecCPU
benchmarks that were impacted the most. Our measurements confirm the authors'
claim that the runtimes of the other C++ benchmarks are virtually unaffected.
The leftmost \piCFI column should be compared to the reported column for \piCFI.
We measured overheads higher than those reported by Niu and Tan.  Both gobmk and
xalancbmk show markedly higher performance overheads in our experiments; we
believe this is in part explained by the fact that Niu and Tan used a newer
Intel Xeon processor having an improved branch
predictor~\cite{rohou.etal+15-branch-predict} and higher clock speeds (3.4 vs
2.2 GHz).

We ran \piCFI in both normal mode and with tail calls disabled. The geometric
mean
overhead increased by 1.9\% with tail calls disabled.  Disabling tail calls in
turn increases the number of equivalence classes on each benchmark
\autoref{fig:equivalence-classes}.  This is a classic example of the
performance/security precision trade-off when designing CFI mechanisms.
Implementers can choose the most precise policy within their performance target.
\CFGuard offers the most efficient protection of forward indirect branches
whereas \piCFI offers higher security at slightly higher cost.

\subsection{Reported CFI Performance}
\label{sec:reported-perf}

The right-hand side of \autoref{tbl:measured-perf} lists reported overheads on
\SpecCPU for CFI mechanisms that we do not measure.  \IFCC is the first CFI
mechanism implemented in LLVM which was later replaced by \LLVMCFI.  \MCFI is
the precursor to \piCFI.  \PathArmor is a recent CFI mechanism that uses
dynamic binary rewriting and a hardware feature, the Last Branch Record
(LBR)~\cite{intelinstructionmanual} register, that traces the 16 most recently
executed indirect control-flow transfers.  \LD is a pure dynamic binary
translation approach to CFI that includes precise enforcement of returns using a
shadow stack.  \CryptoCFI is a compiler-based approach which stores a
cryptographically-secure hash-based message authentication code, HMAC, next to
each pointer.  Checking the HMAC of a pointer before indirect branches avoids a
static points-to analysis to generate a CFG.
\ROPecker is a CFI mechanism that uses a combination of offline analysis, traces
recorded by the LBR register, and emulation in an attempt to detect ROP attacks.
Finally, the \binCFI approach uses static binary rewriting like the original CFI
mechanism; \binCFI is notable for its ability to protect stripped,
position-independent ELF binaries that do not contain relocation information.

The reported overheads match our measurements: xalancbmk and povray impose the
highest overheads---up to 15\% for ROPecker, which otherwise exhibits low
overheads, and 1.7x for \CryptoCFI. The interpreter benchmark, perlbench,
executes a high number of indirect branches, which leads to high overheads,
particularly for \LD, \PathArmor, and \binCFI.

Looking at CFI mechanisms that do not require re-compilation---\PathArmor, \LD,
\ROPecker, and \binCFI we see that the mechanisms that only check the contents
of the LBR before system calls (\PathArmor and \ROPecker) report lower mean
overheads than approaches that comprehensively instrument indirect branches (\LD
and \binCFI) in existing binaries.  More broadly, comparing compiler-based
mechanisms with binary-level mechanisms, we see that compiler-based approaches
are typically as efficient as the binary-level mechanisms that trace control
flows using the LBR although compiler-based mechanisms do not limit
protection to a short window of recently executed branches.  More comprehensive
binary-level mechanisms, \LD and \binCFI generally have higher overheads than
compiler-based equivalents. On the other hand, \LD shows the advantage of binary
translation: almost any program can be analyzed and protected, independent
from the compiler and source code. Also note that \LD incurs additional overhead
for its shadow stack, while none of the other mechanisms
in~\autoref{tbl:measured-perf} have a shadow stack.

Although we cannot directly compare the reported
overheads of \binCFI with our measured overheads for \CFGuard, the
mechanisms enforce CFI policies of roughly similar precision
(compare \autoref{fig:spider:bincfi} and \autoref{fig:spider:cfguard}).
\CFGuard, however, has a substantially lower performance overhead. This is not
surprising given that compilers operate on a high-level program representation
that is more amenable to static program analysis and optimization of the CFI
instrumentation.  On the other hand, compiler-based CFI mechanisms are not
strictly faster than binary-level mechanisms, \CryptoCFI has the highest
reported overheads by far although it is implemented in the LLVM compiler.

\autoref{tbl:external_avg} surveys CFI approaches that do not report overheads
using the \SpecCPU benchmarks like the majority of recent CFI mechanisms do.
Some authors, use an older version of the SPEC benchmarks~\cite{abadi.etal+05,
mohan.etal+15} whereas others evaluate performance using, e.g., web
browsers~\cite{jang.etal+14,CCFIR}, or web servers~\cite{xia.etal+12-cfimon,
lockdown}.
Although it is valuable to quantify overheads of CFI enforcement on more modern
and realistic programs, it remains helpful to include the overheads for \SpecCPU
benchmarks.

\renewcommand{\arraystretch}{1} %

\begin{table*}[th]
  \small
	\centering
	\caption{CFI performance overhead (\%) reported from previous publications. A label of $^C$ indicates we computed the geometric mean overhead over the listed benchmarks, otherwise it is the published average.}
	\label{tbl:external_avg}
  \begin{tabular}{l|l|r}
  & Benchmarks & Overhead \\
  \hline
  \ROPGuard~\cite{ropguard} & PCMark Vantage, NovaBench, 3DMark06, Peacekeeper, & 0.5\% \\
  & Sunspider,  SuperPI 16M & \\
  \SD~\cite{jang.etal+14} & Octane, Kraken, Sunspider, Balls, linelayout, HTML5 & 2.0\% \\
  \CCFIR~\cite{CCFIR} & SPEC2kINT, SPEC2kFP, SPEC2k6INT & $^C$ 2.1\% \\
  \KBouncer~\cite{kBouncer} & wmplayer, Internet Explorer, Adobe Reader & $^C$ 4.0\% \\
  \OCFI~\cite{mohan.etal+15} & SPEC2k & 4.7\% \\
  \CFIMon~\cite{xia.etal+12-cfimon} & httpd, Exim, Wu-ftpd, Memcached & 6.1\% \\
  Original CFI~\cite{abadi.etal+05} & SPEC2k & 16.0\% \\
  \end{tabular}
\end{table*}

\subsection{Discussion}

As \autoref{tbl:measured-perf} shows, authors working in the area of CFI seem
to agree to evaluate their mechanisms using the \SpecCPU benchmarks. There is,
however, less agreement on whether to include both the integer and floating
point subsets. The authors of \LD report the most complete set of benchmark
results covering both integer and floating point benchmarks and the authors of
\binCFI, \piCFI, and \MCFI include most of the integer benchmarks and a subset
of the floating point ones. The authors of \VTV and \IFCC only report subsets of
integer and floating point benchmarks where their solutions introduce
non-negligible overheads.
Except for CFI mechanisms focused on a particular type of control flows such as
virtual method calls, authors should strive to report overheads on the full
suite of \SpecCPU benchmarks. In case there is insufficient time to evaluate a
CFI mechanism on all benchmarks, we strongly encourage authors to focus on the
ones that are challenging to protect with low overheads. These include
perlbench, gcc, gobmk, sjeng, omnetpp, povray, and xalancbmk.  Additionally, it
is desirable to supplement \SpecCPU measurements with measurements for large,
frequently targeted applications such as web browsers and web servers.

Although ``traditional'' CFI mechanisms (e.g., those that check indirect branch
targets using a pre-computed CFG) can be implemented most efficiently in a
compiler, this does not automatically make such solutions superior to binary-level
CFI mechanisms. The advantages of the latter type of approaches include, most
prominently, the ability to work directly on stripped binaries when the
corresponding source is unavailable. This allows CFI enforcement to be applied
independently of the code producer and therefore puts the performance/security
trade off in the hands of the end-users or system administrators.  Moreover,
binary-level solutions naturally operate on the level of entire program modules
irrespective of the source language, compiler, and compilation mode that was
used to generate the code.  Implementers of compiler-based CFI solutions on the
other hand must spend additional effort to support separate compilation or
require LTO operation which, in some instances, lowers the usability of the CFI
mechanism~\cite{szekeres.etal+13}.

\section{Cross-cutting Concerns}\label{sec:xcutting}

This section discusses CFI enforcement mechanisms, presents calls to action
identified by our study for the CFI
community, and identifies current frontiers in CFI research.

\subsection{Enforcement Mechanisms}

The CFI precursor Program Shepherding~\cite{kiriansky02sec} was built on top of
a dynamic optimization engine, RIO.  For CFI like security policies, Program
Shepherding effects the way RIO links basic blocks together on indirect calls.
They improve the performance overhead of this approach by maintaining traces, or
sequences of basic blocks, in which they only have to check that the indirect
branch target is the same.

Many CFI papers follow the ID-based scheme presented by Abadi et.
al~\cite{abadi.etal+05}.  This scheme assigns a label to each indirect control
flow transfer, and to each potential target in the program.  Before the
transfer, they insert instrumentation to insure that the label of the control
flow transfer matches the label of the destination.

Recent work from Google~\cite{tice.etal+14,LLVMCFI} and Microsoft~\cite{CFGUARD}
has moved beyond the ID-based schemes to optimized set checks.  These rely on
aligning metadata such that pointer transformations can be performed quickly
before indirect jumps.  These transformations guarantee that the indirect jump
target is valid.

{\bf Hardware-Supported Enforcement}
Modern processors offer several hardware security-oriented features. Data
Execution Prevention is a classical example of how a simple hardware feature can
eliminate an entire class of attacks. Many processors also support AES
encryption, random number generation, secure enclaves, and array bounds checking
via instruction set extensions.

Researchers have explored architectural support for CFI
enforcement~\cite{davi.etal+14-hw-cfi,arias.etal+15-hafix,sullivan.etal+16-hwcfi,christoCOPASPY16}
with the goal of lowering performance overheads. A particular advantage of these
solutions is that backward edges can be protected by a fully-isolated shadow
stack with an average overhead of just 2\% for protection of forward and
backward edges. This stands in contrast to the average overheads for
software-based shadow stacks which range from  3 to 14\% according
to~\citet{dang-shadow_stacks}.

There have also been efforts to repurpose existing hardware mechanisms to
implement CFI~\cite{kBouncer,ropecker,veen.etal+15,yuan.etal+15}.
\citet{kBouncer} were first to demonstrate a CFI mechanism using the 16-entry
LBR branch trace facility of Intel x86 processors. The key idea in their
kBouncer solution is to check the control flow path that led up to a potentially
dangerous system call by inspecting the LBR; a heuristic was used to distinguish
execution traces induced by ROP chains from legitimate execution traces.
ROPecker by~\citet{ropecker} subsequently extended LBR-based CFI enforcement to
also emulate what code would execute past the system call. While these
approaches offer negligible overheads and do not require recompilation of existing
code, subsequent research showed that carefully crafted ROP attacks can bypass
both of these mechanisms~\cite{out-of-control, ROP-CFI:Davi, ROP-CFI:Wagner}.
The CFIGuard mechanism~\cite{yuan.etal+15} uses the LBR feature in
conjunction with hardware performance counters to heuristically detect ROP
attacks.  \cite{xia.etal+12-cfimon} used the branch trace store, which records
control-flow transfers to a buffer in memory, rather than the LBR for CFI
enforcement.  \citet{mashtizadeh.etal+15}'s \CryptoCFI uses the Intel AES-NI
instruction set to compute cryptographically-enforced hash-based message
authentication codes, HMACs, for pointers stored in attacker-observable memory.
By verifying HMACs before pointers are used, \CryptoCFI prevents control-flow
hijacking.  \citet{mohan.etal+15} leverage Intel's MPX instruction set extension
by re-casting the problem of CFI enforcement as a bounds checking problem over a
randomized CFG.

Most recently, Intel announced hardware support for CFI in future x86
processors~\cite{intel-cet-hw-cfi}.  Intel Control-flow Enforcement Technology
(CET) adds two new instructions, ENDBR32 and ENDBR64, for forward edge protection.
Under CET, the target of any indirect jump or indirect call must be a ENDBR
instruction.  This provides coarse-grained protection where any of the
possible indirect targets are allowed at every indirect control-flow transfer.
There is only one equivalence class which contains every ENDBR instruction in
the program.  For backward edges, CET provides a new Shadow Stack Pointer (SSP)
register which is exclusively manipulated by new shadow stack instructions.
Memory
used by the shadow stack resides in virtual memory and is protected with page
permissions.  In summary, CET provides precise backward edge protection using
a shadow stack, but forward edge protection is imprecise because
there is only one possible label for destinations.

\subsection{Open Problems}

As seen in \autoref{ss:eval-qualitative} most existing CFI implementations use ad
hoc, imprecise analysis techniques when constructing their CFG.  This
unnecessarily weakens these mechanisms, as seen in
\autoref{ss:eval-quantitative}.  All future work in CFI should use
flow-sensitive and
context-sensitive analysis for forward edges, SAP.F.5 from \autoref{sss:sap}.  On
backward edges, we recommend shadow stacks as they have negligible overhead and
are more precise than any possible static analysis.
In this same vein, a study of real world applications that identifies coding
practices that lead to large equivalence classes would be immensely helpful.
This could lead to coding best practices that dramatically increase the security
provided by CFI.

Quantifying the incremental security provided by CFI, or any other security
mechanism, is an open problem.  However, a large adversarial analysis study
would provide additional insight into the security provided by CFI.  Further, it
is likely that CFI could be adapted as a result of such a study to make attacks
more difficult.

\subsection{Research Frontiers}
Recent trends in CFI research target improving CFI in directions beyond new
analysis or enforcement algorithms.
Some approaches have sought to increase CFI protection coverage to include just-in-time code and operating system kernels.
Others leverage advances in hardware to improve performance or enable new enforcement strategies.
We discuss these research directions in the CFI landscape which cross-cut the traditional categories of performance and security.

{\bf Protecting Operating System Kernels.}
In monolithic kernels, all kernel
software is running at the same privilege levels and any memory corruption can
be fatal for security. A kernel is vastly different from a user-space
application as it is directly exposed to the underlying hardware and an attacker
in that space has access to privileged instructions that may change interrupts,
page table structures, page table permissions, or privileged data structures.
KCoFI~\cite{criswell.etal+14} introduces a first CFI policy for commodity
operating systems and considers these specific problems. The CFI mechanism is
fairly coarse-grained: any indirect function call may target any valid functions
and returns may target any call site (instead of executable bytes).
Xinyang Ge et al.~\cite{ge16eurosp} introduce a precise CFI policy
inference mechanism by leveraging common function pointer usage patterns in
kernel code (SAP.F.4b on the forward edge and SAP.B.1 on the backward edge).

{\bf Protecting Just-in-time Compiled Code.}
Like other defenses, it is important that CFI is deployed comprehensively since
adversaries only have to find a single unprotected indirect branch to compromise
the entire process.  Some applications contain just-in-time, JIT, compilers that
dynamically emit machine code for managed languages such as Java and JavaScript.
\citet{niu.tan+14-rockjit} presented RockJIT, a CFI mechanism that
specifically targets the additional attack surface exposed by JIT compilers.
RockJIT faces two challenges unique to dynamically-generated code:
(i)~the code heap used by JIT compilers is usually simultaneously writable and executable to
allow important optimizations such as inline caching~\cite{holzle.ungar+94} and on-stack replacement,
(ii)~computing the control-flow graphs for dynamic languages during execution
without imposing substantial performance overheads.
RockJIT solves the first challenge by replacing the original heap with a shadow code heap which is
readable and writable but not executable and by introducing a sandboxed code
heap which is readable
and executable, but not writable.
To avoid increased memory consumption, RockJIT maps the sandboxed code heap and the shadow heap to
the same physical memory pages with different permissions.
RockJIT addresses the second challenge by both (i) modifying the JIT compiler to emit meta-data about
indirect branches in the generated code and (ii) enforcing a coarse-grained CFI policy on
JITed code which avoids the need for static analysis.
The authors argue that a less precise CFI policy for JITed code is acceptable as
long as both
(i)~the host application is protected by a more precise policy and
(ii)~JIT-compiled code prevents adversaries from making system calls.
In the Edge browser, Microsoft has updated the JIT compilers for JavaScript and
Flash to instrument generated calls and to inform \CFGuard of new control-flow targets
through calls to \texttt{SetProcessValidCallTargets}~\cite{msdn16, bheu15, bh16}.

{\bf Protecting Interpreters.}
Control-flow integrity for interpreters faces similar challenges as just-in-time compilers.
Interpreters are widely deployed, e.g., two major web browsers, Internet
Explorer and Safari, rely on mixed-mode execution models that interpret code
until it becomes ``hot'' enough for just-in-time compilation~\cite{aycock+03},
and some Desktop software, too, is interpreted, e.g., Dropbox's client is implemented in Python.
We have already described the ``worst-case'' interpreters pose to CFI from a security perspective:
even if the interpreter's code is protected by CFI, its actual functionality is determined by a program in data memory.
This separation has two important implications:
(i) static analysis for an interpreter dispatch routine will result in an over-approximation, and
(ii) it enables non-control data attacks through manipulating program source code in writeable data memory prior to JIT compilation.

Interpreters are inherently dynamic, which on the one hand means, CFI for
interpreters could rely on precise dynamic points-to information, but on the
other hand also indicates problems to build a complete control-flow graph for
such programs. Dynamically executing strings as code (\texttt{eval}) further complicates this.
Any CFI mechanism for interpreters needs to address this challenge.

{\bf Protecting Method Dispatch in Object-Oriented Languages.}
In C/\C++ method calls use vtables, which contain addresses to methods, to dynamically bind methods according to the dynamic type of an object.
This mechanism is, however, not the only possible way to implement dynamic binding.
Predating \C++, for example, is Smalltalk-style method dispatch, which influenced the method dispatch mechanisms in other languages, such as Objective-C and JavaScript.
In Smalltalk, all method calls are resolved using a dedicated function called \texttt{send}.
This \texttt{send} function takes two parameters:
(i) the object (also called the receiver of the method call), and
(ii) the method name.
Using these parameters, the \texttt{send} method determines, at call-time, which method to actually invoke.
In general, the determination of which methods are eligible call targets, and which methods cannot be invoked for certain objects and classes cannot be computed statically.
Moreover, since objects and classes are both data, manipulation of data to hijack control-flow suffices to influence the method dispatch for malicious intent.
While Pewny and Holz~\cite{pewny.holz+13} propose a mechanism for Objective-C
send-like dispatch, the generalisation to Smalltalk-style dispatch remains
unsolved.

\section{Conclusions}\label{sec:conclusions}

Control-flow integrity substantially raises the bar against attacks that exploit
memory corruption vulnerabilities to execute arbitrary code.  In the decade
since its inception, researchers have made major advances and explored a great
number of materially different mechanisms and implementation choices.  Comparing
and evaluating these mechanisms is non-trivial and most authors only provide
ad-hoc security and performance evaluations.  A prerequisite to any systematic
evaluation is a set of well-defined metrics.  In this paper, we have proposed
metrics to qualitatively (based on the underlying analysis) and quantitatively
(based on a practical evaluation) assess the security benefits of a
representative sample of CFI mechanisms.  Additionally, we have evaluated
the performance trade-offs and have surveyed cross-cutting concerns and their
impacts on the applicability of CFI.

Our systematization serves as an entry point and guide to the now voluminous and
diverse literature on control-flow integrity.  Most importantly, we capture the
current state of the art in terms of precision and performance. We report
large variations in the forward and backward edge precision for the evaluated
mechanisms with corresponding performance overhead: higher precision results in
(slightly) higher performance overhead.

We hope that our unified nomenclature will gradually displace the ill-defined
qualitative distinction between ``fine-grained'' and ``coarse-grained'' labels
that authors apply inconsistently across publications.  Our metrics provide the
necessary guidance and data to compare CFI implementations in a more nuanced
way. This helps software developers and compiler writers gain appreciation for
the performance/security trade-off between different CFI mechanisms.  For the
security community, this work provides a map of what has been done, and
highlights fertile grounds for future research.  Beyond metrics, our unified
nomenclature allows clear distinctions of mechanisms. These metrics, if adopted,
are useful to evaluate and describe future improvements to CFI.

\section*{Acknowledgments}
We thank the anonymous reviewers, Antonio Barresi, Manuel Costa, Ben Niu, Gang
Tan, and Matt Miller for
their detailed and constructive feedback.
We also thank Priyam Biswas, Hui Peng,
Andrei Homescu,
Nikhil Gupta,
Divya Varshini Agavalam Padmanabhan,
Prabhu Karthikeyan Rajasekaran, and
Roeland Singer-Heinze
for their help with implementation details,
infrastructure, and helpful discussions.
The evaluation in this survey would not have been possible without the
open-source releases of several CFI mechanisms. We thank the corresponding
authors for open-sourcing their implementation prototypes and encourage researchers to
continue to release them.

\bibliographystyle{ACM-Reference-Format-Journals}
\bibliography{references,PL,integrity,infflow,jit}


\begin{thebibliography}{00}


\ifx \showCODEN    \undefined \def \showCODEN     #1{\unskip}     \fi
\ifx \showDOI      \undefined \def \showDOI       #1{{\tt DOI:}\penalty0{#1}\ }
  \fi
\ifx \showISBNx    \undefined \def \showISBNx     #1{\unskip}     \fi
\ifx \showISBNxiii \undefined \def \showISBNxiii  #1{\unskip}     \fi
\ifx \showISSN     \undefined \def \showISSN      #1{\unskip}     \fi
\ifx \showLCCN     \undefined \def \showLCCN      #1{\unskip}     \fi
\ifx \shownote     \undefined \def \shownote      #1{#1}          \fi
\ifx \showarticletitle \undefined \def \showarticletitle #1{#1}   \fi
\ifx \showURL      \undefined \def \showURL       #1{#1}          \fi

\bibitem[\protect\citeauthoryear{Abadi, Budiu, Erlingsson, and Ligatti}{Abadi
  et~al\mbox{.}}{2005a}]%
        {abadi.etal+05}
{Martin Abadi}, {Mihai Budiu}, {{\'U}lfar Erlingsson}, {and} {Jay Ligatti}.
  2005a.
\newblock \showarticletitle{Control-Flow Integrity: Principles,
  Implementations, and Applications}. In {\em ACM Conference on Computer and
  Communications Security (CCS)}.
\newblock


\bibitem[\protect\citeauthoryear{Abadi, Budiu, Erlingsson, and Ligatti}{Abadi
  et~al\mbox{.}}{2005b}]%
        {abadi.etal+05-theory-of-cfi}
{Mart\'{\i}n Abadi}, {Mihai Budiu}, {\'{U}lfar Erlingsson}, {and} {Jay
  Ligatti}. 2005b.
\newblock \showarticletitle{A Theory of Secure Control Flow}. In {\em
  Proceedings of the 7th International Conference on Formal Methods and
  Software Engineering} {\em (ICFEM'05)}.
\newblock


\bibitem[\protect\citeauthoryear{Arias, Davi, Hanreich, Jin, Koeberl, Paul,
  Sadeghi, and Sullivan}{Arias et~al\mbox{.}}{2015}]%
        {arias.etal+15-hafix}
{Orlando Arias}, {Lucas Davi}, {Matthias Hanreich}, {Yier Jin}, {Patrick
  Koeberl}, {Debayan Paul}, {Ahmad-Reza Sadeghi}, {and} {Dean Sullivan}. 2015.
\newblock \showarticletitle{{HAFIX}: Hardware-Assisted Flow Integrity
  Extension}. In {\em Annual Design Automation Conference (DAC)}.
\newblock


\bibitem[\protect\citeauthoryear{Aycock}{Aycock}{2003}]%
        {aycock+03}
{John Aycock}. 2003.
\newblock \showarticletitle{{A brief history of just-in-time}}.
\newblock {\it Comput. Surveys} {35}, 2 (2003), 97--113.
\newblock


\bibitem[\protect\citeauthoryear{Bacon and Sweeney}{Bacon and Sweeney}{1996}]%
        {Bacon1996}
{David~F. Bacon} {and} {Peter~F. Sweeney}. 1996.
\newblock \showarticletitle{{Fast static analysis of C++ virtual function
  calls}}.
\newblock {\em ACM SIGPLAN Notices\/} {31}, 10 (oct 1996), 324--341.
\newblock


\bibitem[\protect\citeauthoryear{Bell}{Bell}{1973}]%
        {bell+73}
{James~R Bell}. 1973.
\newblock \showarticletitle{{Threaded code}}.
\newblock {\it Commun. ACM} {16}, 6 (jun 1973), 370--372.
\newblock
\showISSN{00010782}


\bibitem[\protect\citeauthoryear{Bletsch, Jiang, and Freeh}{Bletsch
  et~al\mbox{.}}{2011}]%
        {Bletsch2011a}
{Tyler Bletsch}, {Xuxian Jiang}, {and} {Vince Freeh}. 2011.
\newblock \showarticletitle{{Mitigating code-reuse attacks with control-flow
  locking}}. In {\em Annual Computer Security Applications Conference (ACSAC)}.
  New York, New York, USA.
\newblock


\bibitem[\protect\citeauthoryear{Bounov, Kici, and Lerner}{Bounov
  et~al\mbox{.}}{2016}]%
        {bounov.etal+15}
{Dimitar Bounov}, {Rami Kici}, {and} {Sorin Lerner}. 2016.
\newblock \showarticletitle{Protecting {C++} Dynamic Dispatch Through VTable
  Interleaving}. In {\em Symposium on Network and Distributed System Security
  (NDSS)}.
\newblock
\newblock
\shownote{To appear.}


\bibitem[\protect\citeauthoryear{Carlini, Barresi, Payer, Wagner, and
  Gross}{Carlini et~al\mbox{.}}{2015}]%
        {cfbending}
{Nicholas Carlini}, {Antonio Barresi}, {Mathias Payer}, {David Wagner}, {and}
  {Thomas~R. Gross}. 2015.
\newblock \showarticletitle{Control-Flow Bending: On the Effectiveness of
  Control-Flow Integrity}. In {\em 24th {USENIX} Security Symposium, {USENIX}
  Security 15, Washington, D.C., USA, August 12-14, 2015.}
\newblock


\bibitem[\protect\citeauthoryear{Carlini and Wagner}{Carlini and
  Wagner}{2014}]%
        {ROP-CFI:Wagner}
{Nicholas Carlini} {and} {David Wagner}. 2014.
\newblock \showarticletitle{{ROP} is Still Dangerous: Breaking Modern
  Defenses}. In {\em USENIX Security Symposium}.
\newblock


\bibitem[\protect\citeauthoryear{Checkoway, Davi, Dmitrienko, Sadeghi, Shacham,
  and Winandy}{Checkoway et~al\mbox{.}}{2010}]%
        {Shacham10}
{Stephen Checkoway}, {Lucas Davi}, {Alexandra Dmitrienko}, {Ahmad{-}Reza
  Sadeghi}, {Hovav Shacham}, {and} {Marcel Winandy}. 2010.
\newblock \showarticletitle{Return-oriented programming without returns}. In
  {\em ACM Conference on Computer and Communications Security (CCS)}.
\newblock


\bibitem[\protect\citeauthoryear{Cheng, Zhou, Miao, Ding, and Deng}{Cheng
  et~al\mbox{.}}{2014}]%
        {ropecker}
{Yueqiang Cheng}, {Zongwei Zhou}, {Yu Miao}, {Xuhua Ding}, {and} {Robert~Huijie
  Deng}. 2014.
\newblock \showarticletitle{{ROPecker}: A Generic and Practical Approach For
  Defending Against {ROP} Attacks}. In {\em Symposium on Network and
  Distributed System Security (NDSS)}.
\newblock


\bibitem[\protect\citeauthoryear{Christoulakis, Christou, Athanasopoulos, and
  Ioannidis}{Christoulakis et~al\mbox{.}}{2016}]%
        {christoCOPASPY16}
{Nick Christoulakis}, {George Christou}, {Elias Athanasopoulos}, {and} {Sotiris
  Ioannidis}. 2016.
\newblock \showarticletitle{{HCFI: Hardware-enforced Control-Flow Integrity}}.
  In {\em CODASPY}.
\newblock


\bibitem[\protect\citeauthoryear{Collingbourne}{Collingbourne}{2015}]%
        {LLVMCFI}
{Peter Collingbourne}. 2015.
\newblock {LLVM} --- Control Flow Integrity.
\newblock   (2015).
\newblock
\newblock
\shownote{\url{http://clang.llvm.org/docs/ControlFlowIntegrity.html}.}


\bibitem[\protect\citeauthoryear{Conti, Crane, Davi, Franz, Larsen, Liebchen,
  Negro, Qunaibit, and Sadeghi}{Conti et~al\mbox{.}}{2015}]%
        {conti.etal+15}
{Mauro Conti}, {Stephen Crane}, {Lucas Davi}, {Michael Franz}, {Per Larsen},
  {Christopher Liebchen}, {Marco Negro}, {Mohaned Qunaibit}, {and} {Ahmad-Reza
  Sadeghi}. 2015.
\newblock \showarticletitle{Losing Control: On the Effectiveness of
  Control-Flow Integrity under Stack Attacks}. In {\em ACM Conference on
  Computer and Communications Security (CCS)}.
\newblock


\bibitem[\protect\citeauthoryear{Criswell, Dautenhahn, and Adve}{Criswell
  et~al\mbox{.}}{2014a}]%
        {criswell.etal+14}
{John Criswell}, {Nathan Dautenhahn}, {and} {Vikram Adve}. 2014a.
\newblock \showarticletitle{{KCoFI}: Complete Control-Flow Integrity for
  Commodity Operating System Kernels}. In {\em IEEE Symposium on Security and
  Privacy (S\&P)}.
\newblock


\bibitem[\protect\citeauthoryear{Criswell, Dautenhahn, and Adve}{Criswell
  et~al\mbox{.}}{2014b}]%
        {Criswell2014}
{John Criswell}, {Nathan Dautenhahn}, {and} {Vikram Adve}. 2014b.
\newblock \showarticletitle{{KCoFI: Complete Control-Flow Integrity for
  Commodity Operating System Kernels}}. In {\em 2014 IEEE Symposium on Security
  and Privacy}.
\newblock


\bibitem[\protect\citeauthoryear{Dang, Maniatis, and Wagner}{Dang
  et~al\mbox{.}}{2015}]%
        {dang-shadow_stacks}
{Thurston~H.Y. Dang}, {Petros Maniatis}, {and} {David Wagner}. 2015.
\newblock \showarticletitle{The Performance Cost of Shadow Stacks and Stack
  Canaries}. In {\em ACM Symposium on Information, Computer and Communications
  Security (ASIACCS)}.
\newblock


\bibitem[\protect\citeauthoryear{Davi, Dmitrienko, Egele, Fischer, Holz, Hund,
  N\"urnberger, and Sadeghi}{Davi et~al\mbox{.}}{2012}]%
        {davi.etal+12}
{Lucas Davi}, {Alexandra Dmitrienko}, {Manuel Egele}, {Thomas Fischer},
  {Thorsten Holz}, {Ralf Hund}, {Stefan N\"urnberger}, {and} {Ahmad-Reza
  Sadeghi}. 2012.
\newblock \showarticletitle{{MoCFI}: A Framework to Mitigate Control-Flow
  Attacks on Smartphones}. In {\em Symposium on Network and Distributed System
  Security (NDSS)}.
\newblock


\bibitem[\protect\citeauthoryear{Davi, Koeberl, and Sadeghi}{Davi
  et~al\mbox{.}}{2014a}]%
        {davi.etal+14-hw-cfi}
{Lucas Davi}, {Patrick Koeberl}, {and} {Ahmad-Reza Sadeghi}. 2014a.
\newblock \showarticletitle{Hardware-Assisted Fine-Grained Control-Flow
  Integrity: Towards Efficient Protection of Embedded Systems Against Software
  Exploitation}. In {\em Annual Design Automation Conference (DAC)}.
\newblock


\bibitem[\protect\citeauthoryear{Davi, Lehmann, Sadeghi, and Monrose}{Davi
  et~al\mbox{.}}{2014b}]%
        {ROP-CFI:Davi}
{Lucas Davi}, {Daniel Lehmann}, {Ahmad-Reza Sadeghi}, {and} {Fabian Monrose}.
  2014b.
\newblock \showarticletitle{Stitching the Gadgets: On the Ineffectiveness of
  Coarse-Grained Control-Flow Integrity Protection}. In {\em USENIX Security
  Symposium}.
\newblock


\bibitem[\protect\citeauthoryear{Dean, Grove, and Chambers}{Dean
  et~al\mbox{.}}{1995}]%
        {Dean1995}
{Jeffrey Dean}, {David Grove}, {and} {Craig Chambers}. 1995.
\newblock \showarticletitle{{Optimization of Object-Oriented Programs Using
  Static Class Hierarchy Analysis}}. In {\em European Conference on
  Object-Oriented Programming (ECOOP)}.
\newblock


\bibitem[\protect\citeauthoryear{Debaere and van Campenhout}{Debaere and van
  Campenhout}{1990}]%
        {debaere.campenhout+90}
{Eddy~H Debaere} {and} {Jan~M van Campenhout}. 1990.
\newblock {\em {Interpretation and instruction path coprocessing}}.
\newblock MIT Press.
\newblock
\showISBNx{978-0-262-04107-2}


\bibitem[\protect\citeauthoryear{Evans, Fingeret, Gonzalez, Otgonbaatar, Tang,
  Shrobe, Sidiroglou-Douskos, Rinard, and Okhravi}{Evans
  et~al\mbox{.}}{2015a}]%
        {evans.etal+15}
{Isaac Evans}, {Samuel Fingeret}, {Julian Gonzalez}, {Ulziibayar Otgonbaatar},
  {Tiffany Tang}, {Howard Shrobe}, {Stelios Sidiroglou-Douskos}, {Martin
  Rinard}, {and} {Hamed Okhravi}. 2015a.
\newblock \showarticletitle{Missing the Point: On the Effectiveness of Code
  Pointer Integrity}. In {\em IEEE Symposium on Security and Privacy (S\&P)}.
\newblock


\bibitem[\protect\citeauthoryear{Evans, Long, Otgonbaatar, Shrobe, Rinard,
  Okhravi, and Sidiroglou-Douskos}{Evans et~al\mbox{.}}{2015b}]%
        {Evans15Jujutsu}
{Isaac Evans}, {Fan Long}, {Ulziibayar Otgonbaatar}, {Howard Shrobe}, {Martin
  Rinard}, {Hamed Okhravi}, {and} {Stelios Sidiroglou-Douskos}. 2015b.
\newblock \showarticletitle{Control Jujutsu: On the Weaknesses of Fine-Grained
  Control Flow Integrity}. In {\em Proceedings of the 22Nd ACM SIGSAC
  Conference on Computer and Communications Security}.
\newblock


\bibitem[\protect\citeauthoryear{Falcon}{Falcon}{2015}]%
        {bheu15}
{Francisco Falcon}. 2015.
\newblock Exploiting Adobe Flash Player in the era of Control Flow Guard.
\newblock BlackHat EU'15
  \url{https://www.blackhat.com/docs/eu-15/materials/eu-15-Falcon-Exploiting-Adobe-Flash-Player-In-The-Era-Of-Control-Flow-Guard.pdf}.
    (2015).
\newblock


\bibitem[\protect\citeauthoryear{Fratric}{Fratric}{2012}]%
        {ropguard}
{Ivan Fratric}. 2012.
\newblock {ROPGuard}: Runtime Prevention of Return-Oriented Programming
  Attacks.
\newblock \\\url{http://www.ieee.hr/_download/repository/Ivan_Fratric.pdf}.
  (2012).
\newblock


\bibitem[\protect\citeauthoryear{Gawlik and Holz}{Gawlik and Holz}{2014}]%
        {t-vip}
{Robert Gawlik} {and} {Thorsten Holz}. 2014.
\newblock \showarticletitle{{Towards Automated Integrity Protection of {C++}
  Virtual Function Tables in Binary Programs}}. In {\em Annual Computer
  Security Applications Conference (ACSAC)}.
\newblock


\bibitem[\protect\citeauthoryear{Ge, Talele, Payer, and Jaeger}{Ge
  et~al\mbox{.}}{2016}]%
        {ge16eurosp}
{Xinyang Ge}, {Nirupama Talele}, {Mathias Payer}, {and} {Trent Jaeger}. 2016.
\newblock \showarticletitle{{Fine-Grained Control-Flow Integrity for Kernel
  Software}}. In {\em IEEE European Symp. on Security and Privacy}.
\newblock


\bibitem[\protect\citeauthoryear{G\"oktas, Athanasopoulos, Bos, and
  Portokalidis}{G\"oktas et~al\mbox{.}}{2014}]%
        {out-of-control}
{Enes G\"oktas}, {Elias Athanasopoulos}, {Herbert Bos}, {and} {Georgios
  Portokalidis}. 2014.
\newblock \showarticletitle{Out Of Control: Overcoming Control-Flow Integrity}.
  In {\em IEEE Symposium on Security and Privacy (S\&P)}.
\newblock


\bibitem[\protect\citeauthoryear{Grove and Chambers}{Grove and
  Chambers}{2001}]%
        {Grove2001}
{David Grove} {and} {Craig Chambers}. 2001.
\newblock \showarticletitle{{A framework for call graph construction
  algorithms}}.
\newblock {\em ACM Transactions on Programming Languages and Systems\/} {23}, 6
  (nov 2001), 685--746.
\newblock


\bibitem[\protect\citeauthoryear{Hackett and Aiken}{Hackett and Aiken}{2006}]%
        {Hackett2006}
{Brian Hackett} {and} {Alex Aiken}. 2006.
\newblock \showarticletitle{How is Aliasing Used in Systems Software?}
\newblock {\em Proceedings of the 14th ACM SIGSOFT International Symposium on
  Foundations of Software Engineering\/} (2006), 69--80.
\newblock


\bibitem[\protect\citeauthoryear{Hardekopf and Lin}{Hardekopf and Lin}{2007}]%
        {Hardekopf2007}
{Ben Hardekopf} {and} {Calvin Lin}. 2007.
\newblock \showarticletitle{{The ant and the grasshopper}}. In {\em Proceedings
  of the 2007 ACM SIGPLAN conference on Programming language design and
  implementation - PLDI '07}, Vol.~42. ACM Press, New York, New York, USA, 290.
\newblock
\showISBNx{9781595936332}
\showDOI{%
\url{http://dx.doi.org/10.1145/1250734.1250767}}


\bibitem[\protect\citeauthoryear{Hardekopf and Lin}{Hardekopf and Lin}{2011}]%
        {Hardekopf2011}
{Ben Hardekopf} {and} {Calvin Lin}. 2011.
\newblock \showarticletitle{{Flow-sensitive pointer analysis for millions of
  lines of code}}. In {\em International Symposium on Code Generation and
  Optimization (CGO 2011)}. IEEE, 289--298.
\newblock
\showISBNx{978-1-61284-356-8}
\showISSN{15232867}
\showDOI{%
\url{http://dx.doi.org/10.1109/CGO.2011.5764696}}


\bibitem[\protect\citeauthoryear{Hind}{Hind}{2001}]%
        {Hind2001}
{Michael Hind}. 2001.
\newblock \showarticletitle{{Pointer analysis}}. In {\em Proceedings of the
  2001 ACM SIGPLAN-SIGSOFT workshop on Program analysis for software tools and
  engineering - PASTE '01}. ACM Press, New York, New York, USA, 54--61.
\newblock
\showISBNx{1581134134}
\showDOI{%
\url{http://dx.doi.org/10.1145/379605.379665}}


\bibitem[\protect\citeauthoryear{Hind and Pioli}{Hind and Pioli}{2000}]%
        {Hind2000}
{Michael Hind} {and} {Anthony Pioli}. 2000.
\newblock \showarticletitle{{Which pointer analysis should I use?}}
\newblock {\em ACM SIGSOFT Software Engineering Notes\/} {25}, 5 (sep 2000),
  113--123.
\newblock


\bibitem[\protect\citeauthoryear{H{\"{o}}lzle and Ungar}{H{\"{o}}lzle and
  Ungar}{1994}]%
        {holzle.ungar+94}
{Urs H{\"{o}}lzle} {and} {David Ungar}. 1994.
\newblock \showarticletitle{{Optimizing dynamically-dispatched calls with
  run-time type feedback}}. In {\em ACM SIGPLAN Conference on Programming
  Language Design and Implementation (PLDI)}.
\newblock


\bibitem[\protect\citeauthoryear{{Intel Inc.}}{{Intel Inc.}}{2013}]%
        {intelinstructionmanual}
{{Intel Inc.}} 2013.
\newblock Intel 64 and {IA-32} Architectures. Software Developer's Manual.
\newblock   (2013).
\newblock


\bibitem[\protect\citeauthoryear{Jang, Tatlock, and Lerner}{Jang
  et~al\mbox{.}}{2014}]%
        {jang.etal+14}
{Dongseok Jang}, {Zachary Tatlock}, {and} {Sorin Lerner}. 2014.
\newblock \showarticletitle{{SAFEDISPATCH}: Securing {C++} Virtual Calls from
  Memory Corruption Attacks}. In {\em Symposium on Network and Distributed
  System Security (NDSS)}.
\newblock


\bibitem[\protect\citeauthoryear{Kiriansky}{Kiriansky}{2013}]%
        {kiriansky03thesis}
{Vladimir Kiriansky}. 2013.
\newblock {\em Secure Execution Environment via Program Shepherding}.
\newblock Master's\ thesis. Massachusetts Institute of Technology.
\newblock


\bibitem[\protect\citeauthoryear{Kiriansky, Bruening, and
  Amarasinghe}{Kiriansky et~al\mbox{.}}{2002}]%
        {kiriansky02sec}
{Vladimir Kiriansky}, {Derek Bruening}, {and} {Saman Amarasinghe}. 2002.
\newblock \showarticletitle{Secure Execution Via Program Shepherding}. In {\em
  USENIX Security Symposium}.
\newblock


\bibitem[\protect\citeauthoryear{Kogge}{Kogge}{1982}]%
        {kogge+82}
{Kogge}. 1982.
\newblock \showarticletitle{{An Architectural Trail to Threaded-Code Systems}}.
\newblock {\em Computer\/} {15}, 3 (mar 1982), 22--32.
\newblock
\showISSN{0018-9162}
\showDOI{%
\url{http://dx.doi.org/10.1109/MC.1982.1653970}}


\bibitem[\protect\citeauthoryear{Larsen, Homescu, Brunthaler, and Franz}{Larsen
  et~al\mbox{.}}{2014}]%
        {larsen.etal+14}
{Per Larsen}, {Andrei Homescu}, {Stefan Brunthaler}, {and} {Michael Franz}.
  2014.
\newblock \showarticletitle{{SoK}: Automated Software Diversity}. In {\em IEEE
  Symposium on Security and Privacy (S\&P)}.
\newblock


\bibitem[\protect\citeauthoryear{Lhot{\'{a}}k and Hendren}{Lhot{\'{a}}k and
  Hendren}{2006}]%
        {Lhotak2006}
{O Lhot{\'{a}}k} {and} {Laurie Hendren}. 2006.
\newblock \showarticletitle{{Context-sensitive points-to analysis: is it worth
  it?}}
\newblock {\em Compiler Construction\/} (2006), 47--64.
\newblock


\bibitem[\protect\citeauthoryear{Mashtizadeh, Bittau, Boneh, and
  Mazi{\`{e}}res}{Mashtizadeh et~al\mbox{.}}{2015}]%
        {mashtizadeh.etal+15}
{Ali~Jos{\'{e}} Mashtizadeh}, {Andrea Bittau}, {Dan Boneh}, {and} {David
  Mazi{\`{e}}res}. 2015.
\newblock \showarticletitle{{CCFI:} Cryptographically Enforced Control Flow
  Integrity}. In {\em ACM Conference on Computer and Communications Security
  (CCS)}.
\newblock


\bibitem[\protect\citeauthoryear{McCarty}{McCarty}{2004}]%
        {mccarty+04}
{Bill McCarty}. 2004.
\newblock {\em SELinux: NSA's Open Source Security Enhanced Linux}.
\newblock O'Reilly Media, Inc.
\newblock
\showISBNx{0596007167}


\bibitem[\protect\citeauthoryear{Microsoft}{Microsoft}{2006}]%
        {DEP}
{Microsoft}. 2006.
\newblock {Data Execution Prevention (DEP)}.
\newblock \url{http://support.microsoft.com/kb/875352/EN-US/}.   (2006).
\newblock


\bibitem[\protect\citeauthoryear{Microsoft}{Microsoft}{2015}]%
        {CFGUARD}
{Microsoft}. 2015.
\newblock Visual Studio 2015 --- Compiler Options --- Enable Control Flow
  Guard.
\newblock   (2015).
\newblock
\newblock
\shownote{\url{https://msdn.microsoft.com/en-us/library/dn919635.aspx}.}


\bibitem[\protect\citeauthoryear{Milanova, Rountev, and Ryder}{Milanova
  et~al\mbox{.}}{2002}]%
        {Milanova2002}
{Ana Milanova}, {Atanas Rountev}, {and} {Barbara~G. Ryder}. 2002.
\newblock \showarticletitle{{Parameterized object sensitivity for points-to and
  side-effect analyses for Java}}.
\newblock {\em ACM SIGSOFT Software Engineering Notes\/} {27}, 4 (2002), 1.
\newblock


\bibitem[\protect\citeauthoryear{Miscosoft}{Miscosoft}{2015}]%
        {msdn16}
{Miscosoft}. 2015.
\newblock SetProcessValidCallTargets function.
\newblock
  \url{https://msdn.microsoft.com/en-us/enus/library/windows/desktop/dn934202(v=vs.85).aspx}.
    (2015).
\newblock


\bibitem[\protect\citeauthoryear{Mock, Das, Chambers, and Eggers}{Mock
  et~al\mbox{.}}{2001}]%
        {Mock2001}
{Markus Mock}, {Manuvir Das}, {Craig Chambers}, {and} {Susan~J. Eggers}. 2001.
\newblock \showarticletitle{{Dynamic points-to sets}}. In {\em ACM
  SIGPLAN-SIGSOFT workshop on Program analysis for software tools and
  engineering (PASTE)}.
\newblock


\bibitem[\protect\citeauthoryear{Mohan, Larsen, Brunthaler, Hamlen, and
  Franz}{Mohan et~al\mbox{.}}{2015}]%
        {mohan.etal+15}
{Vishwath Mohan}, {Per Larsen}, {Stefan Brunthaler}, {Kevin Hamlen}, {and}
  {Michael Franz}. 2015.
\newblock \showarticletitle{Opaque Control-Flow Integrity}. In {\em Symposium
  on Network and Distributed System Security (NDSS)}.
\newblock


\bibitem[\protect\citeauthoryear{Nagarakatte, Zhao, Martin, and
  Zdancewic}{Nagarakatte et~al\mbox{.}}{2009}]%
        {softbound}
{Santosh Nagarakatte}, {Jianzhou Zhao}, {Milo~M.K. Martin}, {and} {Steve
  Zdancewic}. 2009.
\newblock \showarticletitle{{SoftBound}: Highly Compatible and Complete Spatial
  Memory Safety for {C}}. In {\em ACM SIGPLAN Conference on Programming
  Language Design and Implementation (PLDI)}.
\newblock


\bibitem[\protect\citeauthoryear{Nagarakatte, Zhao, Martin, and
  Zdancewic}{Nagarakatte et~al\mbox{.}}{2010}]%
        {nagarakatte2010ismm}
{Santosh Nagarakatte}, {Jianzhou Zhao}, {Milo~M.K. Martin}, {and} {Steve
  Zdancewic}. 2010.
\newblock \showarticletitle{{CETS}: Compiler Enforced Temporal Safety for {C}}.
  In {\em ISMM'10}.
\newblock


\bibitem[\protect\citeauthoryear{Nielson, Nielson, and Hankin}{Nielson
  et~al\mbox{.}}{1999}]%
        {Nielson1999}
{Flemming Nielson}, {Hanne~Riis Nielson}, {and} {Chris Hankin}. 1999.
\newblock {\em {Principles of Program Analysis}}.
\newblock Springer Berlin Heidelberg, Berlin, Heidelberg.
\newblock
\showISBNx{978-3-642-08474-4}
\showDOI{%
\url{http://dx.doi.org/10.1007/978-3-662-03811-6}}


\bibitem[\protect\citeauthoryear{Nielson, Nielson, and Hankin}{Nielson
  et~al\mbox{.}}{2009}]%
        {nielson09springer}
{Flemming Nielson}, {Hanne~R. Nielson}, {and} {Chris Hankin}. 2009.
\newblock {\em Principles of Program Analysis}.
\newblock Springer Publishing Company, Incorporated.
\newblock
\showISBNx{3642084745, 9783642084744}


\bibitem[\protect\citeauthoryear{Niu and Tan}{Niu and Tan}{2013}]%
        {Niu2013a}
{Ben Niu} {and} {Gang Tan}. 2013.
\newblock \showarticletitle{{Monitor integrity protection with space efficiency
  and separate compilation}}. In {\em ACM Conference on Computer and
  Communications Security (CCS)}.
\newblock


\bibitem[\protect\citeauthoryear{Niu and Tan}{Niu and Tan}{2014a}]%
        {niu.tan+14}
{Ben Niu} {and} {Gang Tan}. 2014a.
\newblock \showarticletitle{Modular Control-flow Integrity}. In {\em ACM
  SIGPLAN Conference on Programming Language Design and Implementation (PLDI)}.
\newblock


\bibitem[\protect\citeauthoryear{Niu and Tan}{Niu and Tan}{2014b}]%
        {niu.tan+14-rockjit}
{Ben Niu} {and} {Gang Tan}. 2014b.
\newblock \showarticletitle{{RockJIT}: Securing Just-In-Time Compilation Using
  Modular Control-Flow Integrity}. In {\em ACM Conference on Computer and
  Communications Security (CCS)}.
\newblock


\bibitem[\protect\citeauthoryear{Niu and Tan}{Niu and Tan}{2015a}]%
        {the-github-readme-for-mcfi}
{Ben Niu} {and} {Gang Tan}. 2015a.
\newblock MCFI readme.
\newblock \url{https://github.com/mcfi/MCFI/blob/master/README.md}.   (2015).
\newblock


\bibitem[\protect\citeauthoryear{Niu and Tan}{Niu and Tan}{2015b}]%
        {niu.tan+15}
{Ben Niu} {and} {Gang Tan}. 2015b.
\newblock \showarticletitle{Per-Input Control-Flow Integrity}. In {\em
  Proceedings of the 22nd {ACM} {SIGSAC} Conference on Computer and
  Communications Security, Denver, CO, USA, October 12-6, 2015}.
\newblock


\bibitem[\protect\citeauthoryear{Pappas, Polychronakis, and Keromytis}{Pappas
  et~al\mbox{.}}{2013}]%
        {kBouncer}
{Vasilis Pappas}, {Michalis Polychronakis}, {and} {Angelos~D. Keromytis}. 2013.
\newblock \showarticletitle{Transparent {ROP} Exploit Mitigation Using Indirect
  Branch Tracing}. In {\em USENIX Security Symposium}.
\newblock


\bibitem[\protect\citeauthoryear{Patel}{Patel}{2016}]%
        {intel-cet-hw-cfi}
{Baiju Patel}. 2016.
\newblock Intel releases new technology specifications to protect against ROP
  attacks.
\newblock   (2016).
\newblock
\newblock
\shownote{\url{http://blogs.intel.com/evangelists/2016/06/09/intel-release-new-technology-specifications-protect-rop-attacks/}.}


\bibitem[\protect\citeauthoryear{PaX-Team}{PaX-Team}{2003}]%
        {ASLR}
{PaX-Team}. 2003.
\newblock {PaX} {ASLR} ({A}ddress {S}pace {L}ayout {R}andomization).
\newblock \url{http://pax.grsecurity.net/docs/aslr.txt}.   (2003).
\newblock
\showURL{%
\url{http://pax.grsecurity.net/docs/aslr.txt}}


\bibitem[\protect\citeauthoryear{{PaX-Team}}{{PaX-Team}}{2003}]%
        {PaX}
{{PaX-Team}}. 2003.
\newblock PaX Future.
\newblock \url{https://pax.grsecurity.net/docs/pax-future.txt}.   (2003).
\newblock


\bibitem[\protect\citeauthoryear{Payer, Barresi, and Gross}{Payer
  et~al\mbox{.}}{2015}]%
        {lockdown}
{Mathias Payer}, {Antonio Barresi}, {and} {Thomas~R. Gross}. 2015.
\newblock \showarticletitle{Fine-Grained Control-Flow Integrity Through Binary
  Hardening}. In {\em Detection of Intrusions and Malware, and Vulnerability
  Assessment - 12th International Conference, {DIMVA} 2015, Milan, Italy, July
  9-10, 2015, Proceedings}.
\newblock


\bibitem[\protect\citeauthoryear{Pewny and Holz}{Pewny and Holz}{2013}]%
        {pewny.holz+13}
{Jannik Pewny} {and} {Thorsten Holz}. 2013.
\newblock \showarticletitle{Control-flow Restrictor: Compiler-based {CFI} for
  {iOS}}. In {\em Annual Computer Security Applications Conference (ACSAC)}.
\newblock


\bibitem[\protect\citeauthoryear{Prakash, Hu, and Yin}{Prakash
  et~al\mbox{.}}{2015}]%
        {vfguard}
{Aravind Prakash}, {Xunchao Hu}, {and} {Heng Yin}. 2015.
\newblock \showarticletitle{{vfGuard: Strict Protection for Virtual Function
  Calls in {COTS C++} Binaries}}. In {\em Symposium on Network and Distributed
  System Security (NDSS)}.
\newblock


\bibitem[\protect\citeauthoryear{Roemer, Buchanan, Shacham, and Savage}{Roemer
  et~al\mbox{.}}{2012}]%
        {Roemer2011}
{Ryan Roemer}, {Erik Buchanan}, {Hovav Shacham}, {and} {Stefan Savage}. 2012.
\newblock \showarticletitle{Return-Oriented Programming: Systems, Languages,
  and Applications}.
\newblock {\em {ACM} Transactions on Information System Security\/}  {15}
  (2012).
\newblock


\bibitem[\protect\citeauthoryear{Rohou, Swamy, and Seznec}{Rohou
  et~al\mbox{.}}{2015}]%
        {rohou.etal+15-branch-predict}
{Erven Rohou}, {Bharath~Narasimha Swamy}, {and} {Andr{\'e} Seznec}. 2015.
\newblock \showarticletitle{Branch Prediction and the Performance of
  Interpreters: Don'T Trust Folklore}. In {\em IEEE/ACM International Symposium
  on Code Generation and Optimization (CGO)}.
\newblock


\bibitem[\protect\citeauthoryear{Rountev, Kagan, and Gibas}{Rountev
  et~al\mbox{.}}{2004}]%
        {Rountev2004}
{Atanas Rountev}, {Scott Kagan}, {and} {Michael Gibas}. 2004.
\newblock \showarticletitle{{Evaluating the imprecision of static analysis}}.
  In {\em Proceedings of the ACM-SIGPLAN-SIGSOFT workshop on Program analysis
  for software tools and engineering - PASTE '04}. ACM Press, New York, New
  York, USA, 14.
\newblock
\showISBNx{1581139101}
\showDOI{%
\url{http://dx.doi.org/10.1145/996821.996829}}


\bibitem[\protect\citeauthoryear{Sabelfeld and Myers}{Sabelfeld and
  Myers}{2003}]%
        {Sabelfeld2003}
{Andrei Sabelfeld} {and} {A.C. Myers}. 2003.
\newblock \showarticletitle{{Language-based information-flow security}}.
\newblock {\em IEEE Journal on Selected Areas in Communications\/} {21}, 1 (jan
  2003), 5--19.
\newblock
\showISBNx{0733-8716}
\showISSN{0733-8716}
\showDOI{%
\url{http://dx.doi.org/10.1109/JSAC.2002.806121}}


\bibitem[\protect\citeauthoryear{Schuster, Tendyck, Liebchen, Davi, Sadeghi,
  and Holz}{Schuster et~al\mbox{.}}{2015}]%
        {schuster.etal+15}
{Felix Schuster}, {Thomas Tendyck}, {Christopher Liebchen}, {Lucas Davi},
  {Ahmad-Reza Sadeghi}, {and} {Thorsten Holz}. 2015.
\newblock \showarticletitle{{Counterfeit Object-oriented Programming: On the
  Difficulty of Preventing Code Reuse Attacks in {C++} Applications}}. In {\em
  IEEE Symposium on Security and Privacy (S\&P)}.
\newblock


\bibitem[\protect\citeauthoryear{Shacham}{Shacham}{2007}]%
        {ret2libc}
{Hovav Shacham}. 2007.
\newblock \showarticletitle{The Geometry of Innocent Flesh on the Bone:
  Return-into-libc without Function Calls (on the {x86})}. In {\em CCS'07}.
\newblock


\bibitem[\protect\citeauthoryear{Sharir and Pnueli}{Sharir and Pnueli}{1981}]%
        {Sharir1981}
{Micha Sharir} {and} {Amir Pnueli}. 1981.
\newblock \showarticletitle{{Two approaches to interprocedural data flow
  analysis}}. In {\em Program Flow Analysis}, {Steven~S Muchnick} {and} {Neil~D
  Jones} (Eds.). Prentice Hall.
\newblock


\bibitem[\protect\citeauthoryear{Smaragdakis and Balatsouras}{Smaragdakis and
  Balatsouras}{2015}]%
        {Smaragdakis2015}
{Yannis Smaragdakis} {and} {George Balatsouras}. 2015.
\newblock \showarticletitle{{Pointer Analysis}}.
\newblock {\em Foundations and Trends in Programming Languages\/} {2}, 1
  (2015), 1--69.
\newblock
\showISSN{2325-1107}
\showDOI{%
\url{http://dx.doi.org/10.1561/2500000014}}


\bibitem[\protect\citeauthoryear{Smaragdakis, Bravenboer, and
  Lhot{\'{a}}k}{Smaragdakis et~al\mbox{.}}{2011}]%
        {Smaragdakis2011}
{Yannis Smaragdakis}, {Martin Bravenboer}, {and} {Ondrej Lhot{\'{a}}k}. 2011.
\newblock \showarticletitle{{Pick your contexts well}}.
\newblock {\em ACM SIGPLAN Notices\/} {46}, 1 (jan 2011), 17.
\newblock


\bibitem[\protect\citeauthoryear{Sullivan, Arias, Davi, Larsen, Sadeghi, and
  Jin}{Sullivan et~al\mbox{.}}{2016}]%
        {sullivan.etal+16-hwcfi}
{Dean Sullivan}, {Orlando Arias}, {Lucas Davi}, {Per Larsen}, {Ahmad-Reza
  Sadeghi}, {and} {Yier Jin}. 2016.
\newblock \showarticletitle{Strategy Without Tactics: Policy-Agnostic
  Hardware-Enhanced Control-Flow Integrity}. In {\em Annual Design Automation
  Conference (DAC)}.
\newblock


\bibitem[\protect\citeauthoryear{Szekeres, Payer, Wei, and Song}{Szekeres
  et~al\mbox{.}}{2013}]%
        {szekeres.etal+13}
{Laszlo Szekeres}, {Mathias Payer}, {Tao Wei}, {and} {Dawn Song}. 2013.
\newblock \showarticletitle{{SoK}: Eternal War in Memory}. In {\em IEEE
  Symposium on Security and Privacy (S\&P)}.
\newblock


\bibitem[\protect\citeauthoryear{Tice, Roeder, Collingbourne, Checkoway,
  Erlingsson, Lozano, and Pike}{Tice et~al\mbox{.}}{2014}]%
        {tice.etal+14}
{Caroline Tice}, {Tom Roeder}, {Peter Collingbourne}, {Stephen Checkoway},
  {{\'U}lfar Erlingsson}, {Luis Lozano}, {and} {Geoff Pike}. 2014.
\newblock \showarticletitle{Enforcing Forward-Edge Control-Flow Integrity in
  {GCC} \& {LLVM}}. In {\em USENIX Security Symposium}.
\newblock


\bibitem[\protect\citeauthoryear{Tip and Palsberg}{Tip and Palsberg}{2000}]%
        {Tip2000}
{Frank Tip} {and} {Jens Palsberg}. 2000.
\newblock \showarticletitle{{Scalable propagation-based call graph construction
  algorithms}}.
\newblock {\em ACM SIGPLAN Notices\/} {35}, 10 (oct 2000), 281--293.
\newblock


\bibitem[\protect\citeauthoryear{van~de Ven and Molnar}{van~de Ven and
  Molnar}{2004}]%
        {execshield}
{Arjan van~de Ven} {and} {Ingo Molnar}. 2004.
\newblock Exec Shield.
\newblock \url{https://www.redhat.com/f/pdf/rhel/WHP0006US_Execshield.pdf}.
  (2004).
\newblock


\bibitem[\protect\citeauthoryear{van~der Veen, Andriesse, G\"{o}kta\c{s}, Gras,
  Sambuc, Slowinska, Bos, and Giuffrida}{van~der Veen et~al\mbox{.}}{2015}]%
        {veen.etal+15}
{Victor van~der Veen}, {Dennis Andriesse}, {Enes G\"{o}kta\c{s}}, {Ben Gras},
  {Lionel Sambuc}, {Asia Slowinska}, {Herbert Bos}, {and} {Cristiano
  Giuffrida}. 2015.
\newblock \showarticletitle{{PathArmor}: Practical {ROP} Protection Using
  Context-sensitive {CFI}}. In {\em ACM Conference on Computer and
  Communications Security (CCS)}.
\newblock


\bibitem[\protect\citeauthoryear{Wang and Jiang}{Wang and Jiang}{2010}]%
        {wang10oakland}
{Zhi Wang} {and} {Xuxian Jiang}. 2010.
\newblock \showarticletitle{HyperSafe: A Lightweight Approach to Provide
  Lifetime Hypervisor Control-Flow Integrity}. In {\em IEEE S\&P'10}.
\newblock


\bibitem[\protect\citeauthoryear{Weston and Miller}{Weston and Miller}{2016}]%
        {bh16}
{David Weston} {and} {Matt Miller}. 2016.
\newblock Windows 10 Mitigation Improvements.
\newblock BlackHat'16
  \url{https://www.blackhat.com/docs/us-16/materials/us-16-Weston-Windows-10-Mitigation-Improvements.pdf}.
    (2016).
\newblock


\bibitem[\protect\citeauthoryear{Xia, Liu, Chen, and Zang}{Xia
  et~al\mbox{.}}{2012}]%
        {xia.etal+12-cfimon}
{Yubin Xia}, {Yutao Liu}, {Haibo Chen}, {and} {Binyu Zang}. 2012.
\newblock \showarticletitle{CFIMon: Detecting Violation of Control Flow
  Integrity Using Performance Counters}. In {\em IEEE/IFIP Conference on
  Dependable Systems and Networks (DSN)}.
\newblock


\bibitem[\protect\citeauthoryear{Yuan, Zeng, and Ding}{Yuan
  et~al\mbox{.}}{2015}]%
        {yuan.etal+15}
{Pinghai Yuan}, {Qingkai Zeng}, {and} {Xuhua Ding}. 2015.
\newblock \showarticletitle{Hardware-Assisted Fine-Grained Code-Reuse Attack
  Detection}. In {\em International Symposium on Research in Attacks,
  Intrusions and Defenses (RAID)}.
\newblock


\bibitem[\protect\citeauthoryear{Zhang, Song, Chen, Chen, and Song}{Zhang
  et~al\mbox{.}}{2015}]%
        {vtint}
{Chao Zhang}, {Chengyu Song}, {Kevin~Zhijie Chen}, {Zhaofeng Chen}, {and} {Dawn
  Song}. 2015.
\newblock \showarticletitle{{VTint}: Defending Virtual Function Tables'
  Integrity}. In {\em Symposium on Network and Distributed System Security
  (NDSS)}.
\newblock


\bibitem[\protect\citeauthoryear{Zhang, Wei, Chen, Duan, Szekeres, McCamant,
  Song, and Zou}{Zhang et~al\mbox{.}}{2013}]%
        {CCFIR}
{Chao Zhang}, {Tao Wei}, {Zhaofeng Chen}, {Lei Duan}, {Laszlo Szekeres},
  {Stephen McCamant}, {Dawn Song}, {and} {Wei Zou}. 2013.
\newblock \showarticletitle{Practical Control Flow Integrity \& Randomization
  for Binary Executables}. In {\em IEEE Symposium on Security and Privacy
  (S\&P)}.
\newblock


\bibitem[\protect\citeauthoryear{Zhang and Sekar}{Zhang and Sekar}{2013}]%
        {cfi-cots}
{Mingwei Zhang} {and} {R. Sekar}. 2013.
\newblock \showarticletitle{Control Flow Integrity for {COTS} Binaries}. In
  {\em USENIX Security Symposium}.
\newblock


\end{thebibliography}

\appendix
\setcounter{section}{1}

\section{Prior work on static analysis}\label{app:sa-prior-work}

Static analysis research has attracted significant interest from the research
community.  Following our classification of control-flows in
\autoref{ss:cflow-classification}, we are particularly interested in static
analysis that identifies indirect calls/jump targets.  Researchers refer to this
kind of static analysis as \emph{points-to analysis}.  The wealth of information
and results in points-to analysis goes well beyond the scope of this paper. We
refer the interested reader to Smaragdakis and
Balatsouras~\cite{Smaragdakis2015} and focus our attention on how points-to
analysis affects CFI precision. %

\subsection{A Theoretical Perspective}\label{sss:sa-theory}
Many compiler optimizations benefit from points-to analysis.  As a result,
points-to analysis must be sound at all times and therefore conservatively
over-approximates results.  The program analysis
literature (e.g., \cite{Nielson1999,Hind2000,Hind2001,Smaragdakis2015}) expresses this
conservative aspect as a \emph{may}-analysis: A specific object ``may'' point to
any members of a computed points-to set.

For the purposes of this paper, the following orthogonal dimensions in points-to
analysis affect precision:
\begin{itemize}
\item \emph{flow-sensitive} vs. \emph{flow-insensitive}:
  this dimension states whether an analysis considers control-flow (sensitive)
  or not (insensitive).
\item \emph{context-sensitive} vs. \emph{context-insensitive}:
  this dimension states whether an analysis considers various forms of context
  (sensitive) or not (insensitive).
  The literature further separates the following context information sub-categories:
    (i) call-site sensitive:
    the context includes a function's call-site (e.g., call-strings~\cite{Sharir1981}),
    (ii) object sensitive:
    the context includes the specific receiver object present at a
    call-site~\cite{Milanova2002},
    (iii) type sensitive:
    the context includes type information of functions or objects at a call-site~\cite{Smaragdakis2011}.
\end{itemize}
Both dimensions, context and flow sensitivity, are orthogonal and a points-to
analysis combining both yields higher precision.

\begin{figure*}[t]
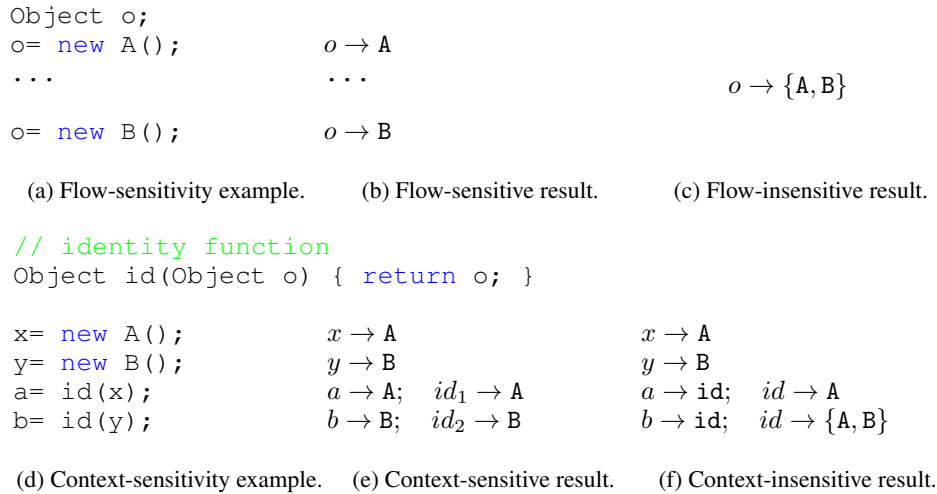

  \centering
  \begin{subfigure}[b]{.3\textwidth}
    \begin{lstlisting}[language=C++,mathescape=true]
Object o;
o= new A();
...

o= new B();
    \end{lstlisting}
    \caption{Flow-sensitivity example.}\label{fig:ex:flow}
  \end{subfigure}%
  \begin{subfigure}[b]{.3\textwidth}
    \begin{lstlisting}[language=C++,mathescape=true]

$o \rightarrow \mathtt{A}$
...

$o \rightarrow \mathtt{B}$
    \end{lstlisting}
    \caption{Flow-sensitive result.}\label{fig:ex:flow-sensitive}
  \end{subfigure}
  \begin{subfigure}[b]{.3\textwidth}
    \begin{eqnarray*}
o \rightarrow \{ \mathtt{A}, \mathtt{B} \}
    \end{eqnarray*}
    \vspace*{1em}
    \caption{Flow-insensitive result.}\label{fig:ex:flow-insensitive}
  \end{subfigure}%
  \\
  \begin{subfigure}[b]{.3\textwidth}
    \begin{lstlisting}[language=C++,mathescape=true]
// identity function
Object id(Object o) { return o; }

x= new A();
y= new B();
a= id(x);
b= id(y);
    \end{lstlisting}
    \caption{Context-sensitivity example.}\label{fig:ex:ctxt}
  \end{subfigure}%
  \begin{subfigure}[b]{.3\textwidth}
    \begin{lstlisting}[language=C++,mathescape=true]


$x \rightarrow \mathtt{A}$
$y \rightarrow \mathtt{B}$
$a \rightarrow \mathtt{A}; \quad id_1 \rightarrow \mathtt{A}$
$b \rightarrow \mathtt{B}; \quad id_2 \rightarrow \mathtt{B}$
    \end{lstlisting}
    \caption{Context-sensitive result.}\label{fig:ex:ctxt-sensitive}
  \end{subfigure}%
  \begin{subfigure}[b]{.3\textwidth}
    \begin{lstlisting}[language=C++,mathescape=true]


$x \rightarrow \mathtt{A}$
$y \rightarrow \mathtt{B}$
$a \rightarrow \mathtt{id}; \quad id \rightarrow \mathtt{A}$
$b \rightarrow \mathtt{id}; \quad id \rightarrow \{ \mathtt{A}, \mathtt{B} \}$
    \end{lstlisting}
    \caption{Context-insensitive result.}\label{fig:ex:ctxt-insensitive}
  \end{subfigure}%
  \caption{Effects of flow/context sensitivity on precision.}\label{fig:ex:context}
  \label{fig:points-to-examples}
\end{figure*}

\paragraph{Flow-Sensitivity}
Figures~\ref{fig:ex:flow} -- \ref{fig:ex:flow-insensitive} show the effect of
flow sensitivity on points-to analysis.  A flow-sensitive analysis considers the
state of the program per line.  We see, for instance,
in~\autoref{fig:ex:flow-sensitive} how a flow-sensitive analysis computes the
proper object type per allocation site.  A flow-insensitive analysis, on the
other hand, computes sets that are valid for the whole program.  Or, simply put,
it lumps all statements of the analyzed block (intra- or interprocedural) into
one set and computes a single points-to set that satisfies all of these
statements.  From a CFI perspective, a flow-sensitive
points-to analysis offers higher precision.
\paragraph{Context-Sensitivity}
Figures~\ref{fig:ex:ctxt} -- \ref{fig:ex:ctxt-insensitive} show the effects of
context sensitivity on points-to analysis.  In \autoref{fig:ex:ctxt} we see that
the function \texttt{id} is called twice, with parameters of different dynamic
types.  Context-insensitive analysis (cf.~\autoref{fig:ex:ctxt-insensitive}),
does not distinguish between the two different calling contexts and therefore
computes an over-approximation by lumping all invocations into one points-to set
(e.g., the result of calling \texttt{id} is a set with two members).  A
context-insensitive analysis, put differently, considers a function independent
from its callers, and is therefore the forward control-flow transfer symmetric
case of a backward control-flow transfers returning to many
callers~\cite{Nielson1999}.
Context-sensitive analysis
(cf.~\autoref{fig:ex:ctxt-sensitive}), on the other hand, uses additional
context information to compute higher precision results.  The last two lines
in~\autoref{fig:ex:ctxt-sensitive} illustrate the higher precision by inferring
the proper dynamic types \texttt{A} and \texttt{B}.  From a CFI perspective, a
context-sensitive points-to analysis offers higher precision.

\lstset{language=C++,
  basicstyle=\ttfamily,
  keywordstyle=\color{blue}\ttfamily,
  stringstyle=\color{red}\ttfamily,
  commentstyle=\color{green}\ttfamily,
  morecomment=[l][\color{magenta}]{\#}
}

\paragraph{Object-Oriented Programming Languages}
A C-like language requires call-string or type context-sensitivity to compute
precise results for function pointers.  Due to dynamic dispatch, however, a
\C++-like language should consider more context provided by object
sensitivity~\cite{Milanova2002,Lhotak2006}.  Alternatively, prior work describes
several algorithms to ``devirtualize'' call-sites.  If a static analysis
identifies that only one receiver is possible for a given call-site (i.e., if
the points-to set is a singleton) a compiler can sidestep expensive dynamic
dispatch via the vtable and generate a direct call to the referenced method.
Class-hierarchy analysis (CHA)~\cite{Dean1995} and rapid-type analysis
(RTA)~\cite{Bacon1996} are prominent examples that use domain-specific
information about the class hierarchy to optimize virtual method calls.  RTA
differs from CHA by pruning entries from the class hierarchy from objects that
have not been instantiated.  As a result, the RTA precision is higher than CHA
precision~\cite{Grove2001}.  Grove and Chambers~\cite{Grove2001} study the topic
of call-graph construction and present a partial order of various
approaches' precision (Figure~19, pg. 735).  With regards to CFI, higher
precision in the call-graph of virtual method invocations translates to either
(i) more de-virtualized call-sites, which replace an indirect call by a
direct call, or (ii) shrinking the points-to sets, which reduce an adversary's
attack surface. Note that the former, de-virtualization of a call-site also has
the added benefit of removing the call-site from a points-to set and
transforming an indirect control-flow transfer to a direct control-flow transfer
that need not be validated by the CFI enforcement component.

\subsection{A Practical Perspective}\label{ss:sa-practical}
Points-to analysis over-approximation reduces precision and therefore restricts
the optimization potential of programs.  The reduced precision also lowers
precision for CFI, opening the door for attackers.
If, for instance, the over-approximated set of computed targets contains many
more ``reachable'' targets, then an attacker can use those control-flow
transfers without violating the CFI policy.  Consequently, prior results from
studying the precision of static points-to analysis are of key importance to
understanding CFI policies' security properties.

Mock et al.~have studied dynamic points-to sets and compared them to statically
determined points-to sets~\cite{Mock2001}.  More precisely, the study used an
instrumentation framework to compute dynamic points-to sets and compared them
with three flow- and context-insensitive points-to algorithms.  The authors
report that static analyses identified 14\% of all points-to sets as
singletons, whereas dynamic points-to sets were singletons in 97\% of all cases.
In addition, the study reports that one out of two statically computed singleton
points-to sets were optimal in the sense that the dynamic points-to sets were
also singletons.  The authors describe some caveats and state that flow and
context sensitive points-to analyses were not practical in evaluation since they
did not scale to practical programs.  Subsequent work has, however, established
the scalability of such points-to
analyses~\cite{Hackett2006,Hardekopf2007,Hardekopf2011}, and a similar
experiment evaluating the precision of computed results is warranted.

Concerning the analysis of devirtualized method calls, prior work reports the
following results.  By way of manual inspection, Rountev et
al.~\cite{Rountev2004} report that 26\% of call chains computed by RTA
were actually infeasible.  Lhotak and Hendren~\cite{Lhotak2006}
studied the effect of context-sensitivity to improve precision on
object-oriented programs.  They find that context sensitivity has only a modest
effect on call-graph precision, but also report substantial benefits of context
sensitivity to resolve virtual calls.  In particular, Lhotak and Hendren
highlight the utility of object-sensitive analyses for this task.  Tip and
Palsberg~\cite{Tip2000} present advanced algorithms, XTA among others, and
report that it improves precision over RTA, on average, by 88\%.

\subsection{Backward Control Flows}\label{sss:backward-cflows}

\begin{figure}[t!]
  \centering
  \begin{tikzpicture}[]
    \node (lblF) at (0, 1) {$f$};
    \node (lblG) at (0, -1) {$g$};
    \node (lblH) at (2, 0) {$h$};
    \node (lblrets) [below right=0.1cm of lblH] {$\{f, g\}$};

    \draw[->] (lblF.east) -- (lblH.north west);
    \draw[->] (lblG.east) -- (lblH.north west);

    \draw[-,dotted] (lblH.north west) -- (lblH.south west);
    \draw[->,dashed] (lblH.south west) -- (lblF.east);
    \draw[->,dashed] (lblH.south west) -- (lblG.east);
  \end{tikzpicture}
  \caption{Backward control-flow precision.
    Solid lines correspond to function calls and dashed lines to returns from functions to their call sites. Call-sites are singletons whereas $h$'s return can return to two callers.}
  \label{fig:backward-cflow}
\end{figure}
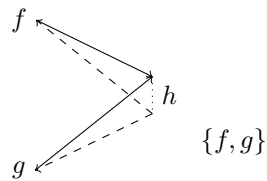

\autoref{fig:backward-cflow} shows two functions, $f$ and $g$, which call
another function $h$.  The return instruction in function $h$ can, therefore,
return to either function $f$ or $g$, depending on which function actually
called $h$ at run-time.  To select the proper caller, the compiler maintains
and uses a stack of activation records, also known as stack frames.  Each stack
frame contains information about the CPU instruction pointer of the caller as
well as bookkeeping information for local variables.

Since there is only one return instruction at the end of a function, even the
most precise static analysis can only infer the set of callers for all calls.
Computing this set, inevitably, leads to imprecision and all call-sites of a
given function must therefore share the same label/ID such that the CFI check
succeeds.  Presently, the only known alternative to this loss of precision is to
maintain a shadow stack and check whether the current return address equals the
return address of the most recent call instruction.

\section{Full quantitative security results}\label{sec:appendix}

\begin{figure*}[hb!]
  \centering
  \includegraphics[scale=.3]{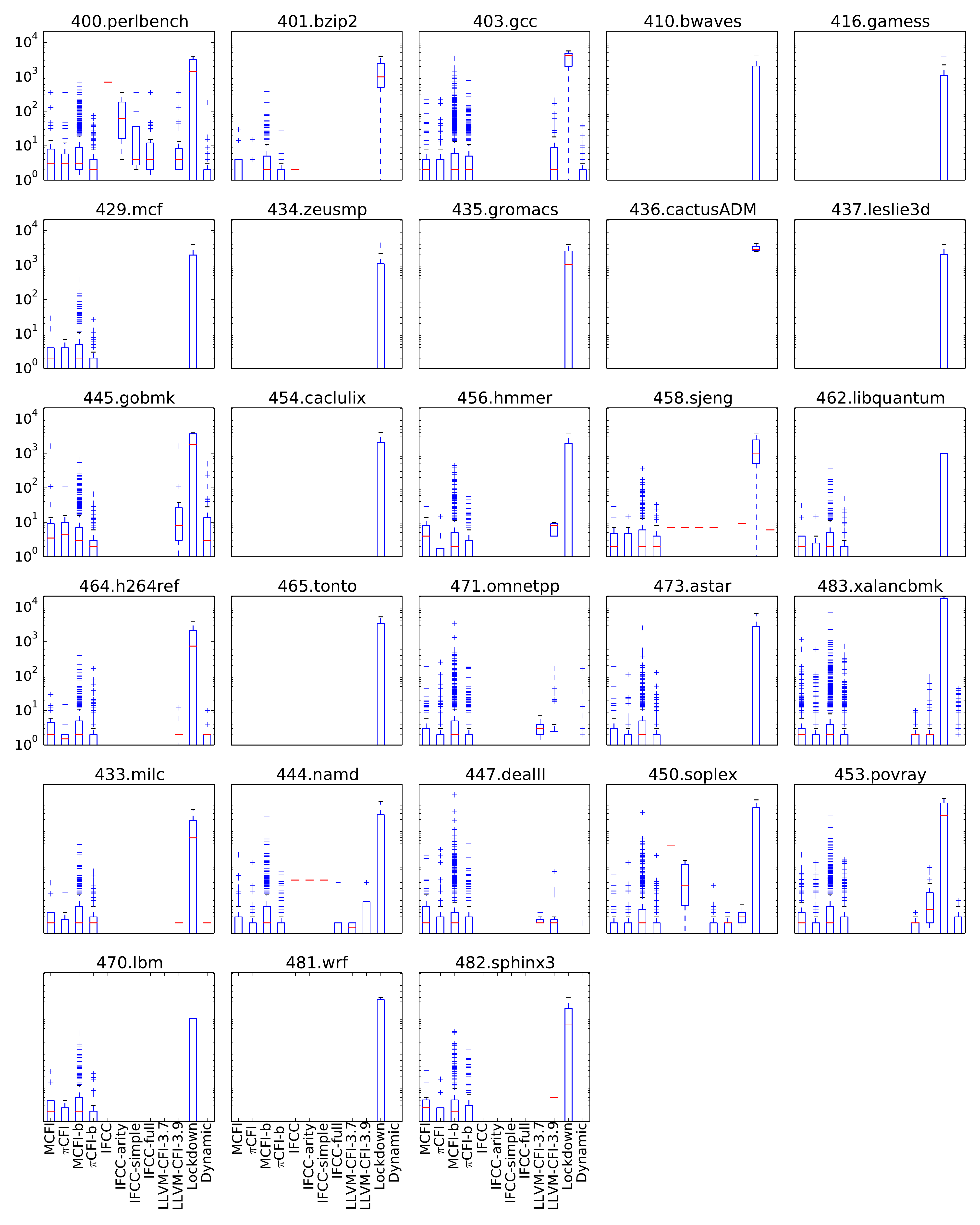}
  \caption{Whisker plot of equivalence classes size for all SPEC CPU2006 benchmarks across
    all implementations (smaller is better).}
  \label{fig:full-whisker}
\end{figure*}

This appendix presents the full quantitative security results.  An abbreviated
version of these results was presented in \autoref{ss:eval-quantitative}.  The full
results are presented here for completeness.  \autoref{tbl:quant-full-results}
contains the number of equivalence sets for each benchmark and every CFI
mechanism that we evaluated.  \autoref{fig:full-whisker} contains the
full set of box and whisker plots.  As this data is fundamentally three
dimensional, these plots are the best way to display it.  As a final note, the
holes in this data reflect the fact that the CFI mechanisms that we evaluated
cannot run the full set of \SpecCPU benchmarks.  This greatly complicates the
task of comparatively evaluating them, as there is only a narrow base of
programs that all the CFI mechanisms run.

\begin{table*}[h!]
	\scriptsize
	\centering
	\begin{tabular}{l|rrrr|rrrr|rrrr|rrrr}
		\toprule
      Benchmark & \multicolumn{15}{c}{CFI Implementation} \\
& \MCFI & \piCFI & \MCFI & \piCFI & \MCFI & \piCFI  & \MCFI & \piCFI &
    \multicolumn{4}{c|}{IFCC} & \multicolumn{2}{c}{\LLVMCFI} & Lock- &
    Dynamic \\

  & \multicolumn{4}{c|}{back edge} & \multicolumn{4}{c|}{forward edge} & single &
    arity & simpl. & full & \multicolumn{1}{c}{3.7} & \multicolumn{1}{c}{3.9}
    & down \\

 & & & \multicolumn{2}{|c|}{no tail call} & & & \multicolumn{2}{|c|}{no tail
  call} & & & & & & & &\\
  \midrule
\gray
400.perlbench & 978 & 310 & 1192 & 429 & 38 & 30 & 38 & 30 & 1 & 6 & 12 & 40 & 0
& 36 & 4 & 83 \\
401.bzip2 & 484 & 82 & 489 & 86 & 14 & 10 & 14 & 10 & 1 & 2 & 2 & 2 & 0 & 2 & 3 & 12 \\
\gray
403.gcc & 2219 & 1260 & 3282 & 1836 & 98 & 90 & 98 & 90 & 0 & 0 & 0 & 0 & 0 & 94
& 3 & 197 \\
429.mcf & 475 & 96 & 475 & 96 & 12 & 8 & 12 & 8 & 0 & 0 & 0 & 0 & 0 & 0 & 3 & 0 \\
\gray
445.gobmk & 922 & 283 & 1075 & 230 & 21 & 17 & 21 & 17 & 0 & 0 & 0 & 0 & 0 & 11
& 4 & 0 \\
456.hmmer & 663 & 134 & 720 & 147 & 14 & 9 & 14 & 9 & 0 & 0 & 0 & 0 & 0 & 3 & 4 & 9 \\
\gray
458.sjeng & 540 & 119 & 557 & 125 & 13 & 9 & 13 & 9 & 1 & 1 & 1 & 1 & 0 & 1 & 3 & 1 \\
462.libquantum & 495 & 88 & 519 & 102 & 12 & 8 & 12 & 8 & 1 & 1 & 1 & 1 & 0 & 0
& 4 & 0 \\
\gray
464.h264ref & 773 & 285 & 847 & 327 & 21 & 15 & 21 & 15 & 0 & 0 & 0 & 0 & 0 & 9
& 4 & 59 \\
471.omnetpp & 1693 & 581 & 1784 & 624 & 357 & 321 & 357 & 321 & 0 & 0 & 0 & 0 &
114 & 35 & 0 & 224 \\
\gray
473.astar & 1096 & 226 & 1108 & 237 & 166 & 150 & 166 & 150 & 0 & 0 & 0 & 0 & 1
& 1 & 6 & 1 \\
483.xalancbmk & 6161 & 2381 & 7162 & 2869 & 1534 & 1200 & 1534 & 1200 & 0 & 0 &
0 & 0 & 2197 & 260 & 6 & 1402 \\
\midrule
\gray
433.milc & 602 & 169 & 628 & 180 & 13 & 9 & 13 & 9 & 0 & 0 & 0 & 0 & 0 & 1 & 4 & 3 \\
444.namd & 1080 & 217 & 1087 & 224 & 166 & 150 & 166 & 150 & 1 & 1 & 1 & 5 & 4
& 4 & 6 & 12 \\
\gray
447.dealII & 2952 & 817 & 3468 & 896 & 293 & 258 & 293 & 258 & 0 & 0 & 0 & 0 &
43 & 15 & 0 & 95 \\
450.soplex & 1444 & 432 & 1569 & 479 & 321 & 291 & 321 & 291 & 1 & 7 & 0 & 186 &
41 & 9 & 6 & 157 \\
\gray
453.povray & 1748 & 650 & 1934 & 743 & 218 & 204 & 218 & 204 & 0 & 0 & 0 & 0 &
29 & 33 & 6 & 49 \\
470.lbm & 465 & 70 & 470 & 74 & 12 & 8 & 12 & 8 & 0 & 0 & 0 & 0 & 0 & 0 & 4 & 0 \\
\gray
482.sphinx3 & 633 & 239 & 677 & 257 & 13 & 9 & 13 & 9 & 0 & 0 & 0 & 0 & 0 & 1 & 4 & 2 \\
      \bottomrule
	\end{tabular}
	\caption{Full quantitative security results for number of equivalence classes.}
	\label{tbl:quant-full-results}
\end{table*}

\end{document}

